\documentclass[sigconf]{acmart}

\usepackage[english]{babel}
\usepackage{blindtext}

\usepackage{graphicx}
\usepackage{multirow}
\usepackage{subcaption}
\usepackage{tabularx}
\usepackage{booktabs}
\usepackage{pifont}  
\renewcommand\footnotetextcopyrightpermission[1]{} % removes footnote with conference info
\setcopyright{none}
\acmDOI{}

\acmISBN{}

\acmConference{arxiv}
\acmYear{2026}
\acmPrice{}

\begin{document}
\def \modelname {ZipCCL}
\title[\modelname{}]{\modelname: Efficient Lossless Data Compression of Communication Collectives for Accelerating LLM Training}

%\titlenote{Produces the permission block, and copyright information}
%\subtitle{Extended Abstract}
\author{
Wenxiang~lin$^{1}$, Xinglin~Pan$^{2}$, Ruibo~Fan$^{2}$,   Shaohuai~Shi$^{1,^\ast}$,  Xiaowen~Chu$^{2,3,^\ast}$}
\affiliation{$^{1}$Harbin Institute of Technology, Shenzhen \country{China}}
\affiliation{$^{2}$The Hong Kong University of Science and Technology (Guangzhou) \country{China}} 
\affiliation{$^{3}$The Hong Kong University of Science and Technology \country{Hong Kong SAR}}
\email{wenxianglin@stu.hit.edu.cn,{xpan413,ruibo.fan   }@connect.hkust-gz.edu.cn}
\email{shaohuais@hit.edu.cn,  xwchu@hkust-gz.edu.cn}
\thanks{$^\ast$Corresponding author.}
% \author{Paper \# 1434, 12 pages body, 14 pages total}
% \author{Firstname Lastname}
% \authornote{Note}
% \affiliation{%
%   \institution{Affiliation}
%   \streetaddress{Address}
%   \city{City} 
%   \state{State} 
%   \postcode{Zipcode}
% }
% \email{email@domain.com}
\renewcommand{\shortauthors}{X.et al.}

\begin{abstract}
Communication has emerged as a critical bottleneck in the distributed training of large language models (LLMs). While numerous approaches have been proposed to reduce communication overhead, the potential of lossless compression has remained largely underexplored since compression and decompression typically consume larger overheads than the benefits of reduced communication traffic. 
% Training data in LLMs—such as activations, gradients, and parameters, often follows a near‑Gaussian distribution, presenting significant compression potential. 
% However, existing GPU‑based lossless compression methods are often prohibitively slow in practice with GPU. 
We observe that the communication data, including activations, gradients and parameters, during training often follows a near‑Gaussian distribution, which is a key feature for data compression. 
Thus, we introduce \modelname{}, a lossless compressed communication library of collectives for LLM training. \modelname{} is equipped with our novel techniques: (1) theoretically ground- ed exponent coding that exploits the Gaussian distribution of LLM tensors to accelerate compression without expensive online statistics, (2) GPU-optimized compression and decompression kernels that carefully design memory access patterns and pipeline using communication-aware data layout, and (3) adaptive communication strategies that dynamically switch collective operations based on workload patterns and system characteristics. 
Evaluated on a 64‑GPU cluster using both mixture-of-experts and dense transformer models, \modelname{} reduces communication time by up to 1.35$\times$ and achieves end‑to‑end training speedups of up to 1.18$\times$ without any impact on model quality.
% while maintaining full precision and convergence integrity. 
% Our work demonstrates that co‑designing compression with communication collectives can unlock significant performance gains in large‑scale LLM training.
\end{abstract}

\maketitle

\section{Introduction}

As large language models (LLMs) continue to scale, training on massive GPU clusters has become inevitable~\cite{liu2024deepseek,qwen3}. To fully utilize these resources, various parallelism strategies have been developed, including Data Parallelism (DP)~\cite{DBLP:dean2012large}, Pipeline Parallelism (PP)~\cite{narayanan2021efficient}, Tensor Parallelism (TP)~\cite{DBLP:dean2012large,narayanan2021efficient}, Expert Parallelism (EP)~\cite{DBLP:Shazeer2017outrageously, hwang2023tutel}, and Fully Sharded Data Parallelism (FSDP)~\cite{zhao2023pytorchfsdp}. While enabling parallel computation across thousands of devices, these methods introduce substantial communication overhead, which often becomes the bottleneck for system scalability.

Existing efforts to mitigate this overhead fall broadly into three categories. The first category focuses on \textbf{scheduling optimization}, which overlaps computation with communication to mask latency~\cite{shi2023pipemoe,lin2025mast,pan2025fsmoe,jiang2024lancet,wang2025spmoe}. The second category involves \textbf{topology-aware operations}, optimizing collectives based on specific network hardware properties~\cite{lin2025hiermoe,liu2024deepseek,blink}. The third category adopts \textbf{communication compression}, primarily utilizing lossy techniques such as quantization~\cite{shi2024schemoe} or sparsification~\cite{Jin_Wang_Zhu_Zhan_Bai_Zhang_Ming_Li_2025}. While effective in reducing data volume, these methods introduce irreversible information loss. For the expensive and complex training of large-scale LLMs, this lossy nature poses significant risks and quantification difficulties regarding model convergence and final accuracy.

\begin{figure}[!t]
   \begin{subfigure}[b]{0.23\textwidth}
		\centering
			\includegraphics[width=\textwidth]{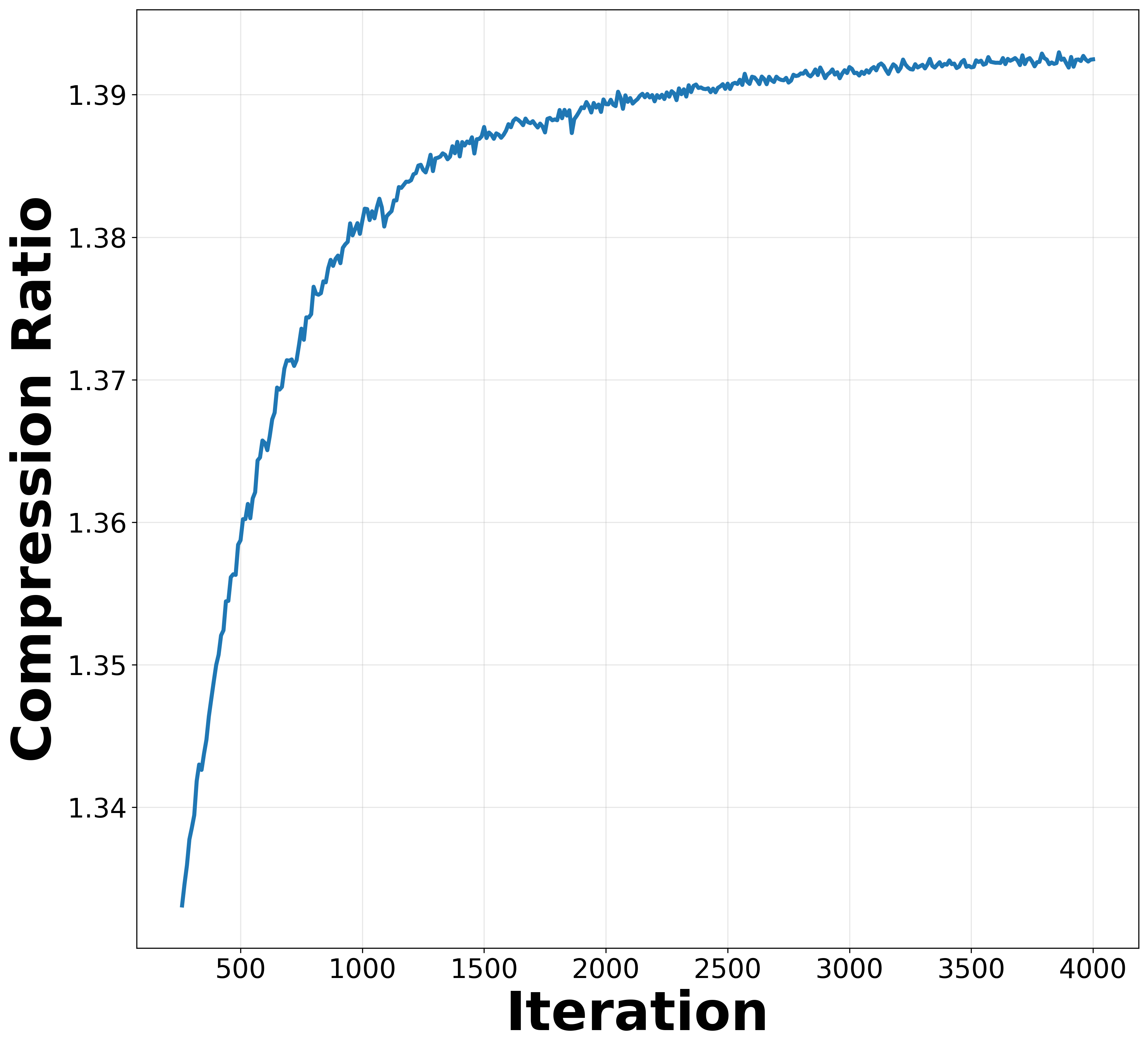}
	\caption{Compression ratio. }
		\label{fig:comp_ratio}
	\end{subfigure}
    	 \begin{subfigure}[b]{0.23\textwidth}
		\centering
		\includegraphics[width=\textwidth]{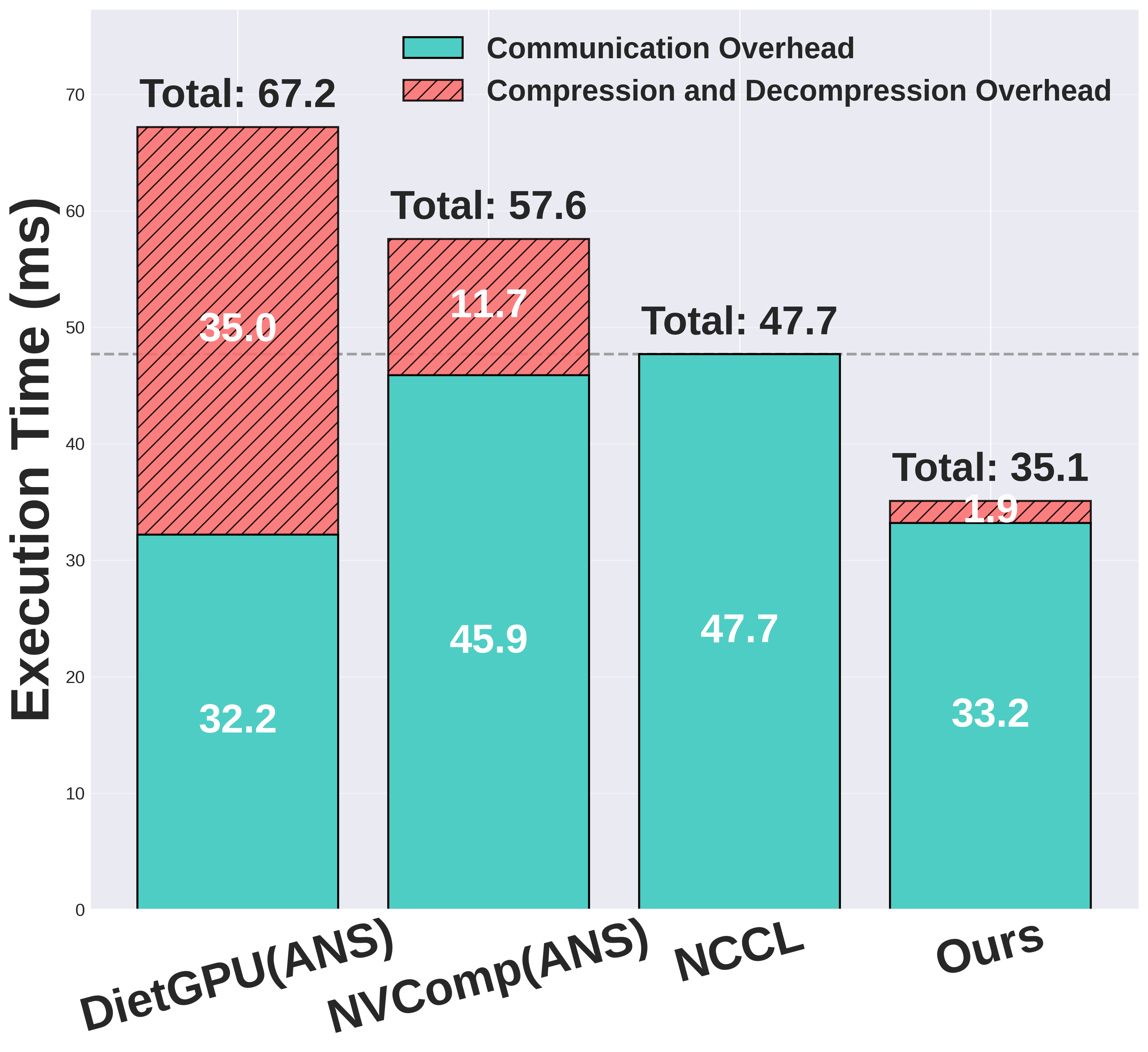}
		\caption{Execution time.}
		\label{fig:compress_compare}
	\end{subfigure}
	\label{fig:comp_rc}
    \caption{Left: Compression ratio (original communication volume / compressed communication volume) averaged over training iterations; Right: Execution time of a lossless compressed All-Gather communication with different lossless compression methods for Llama3-8B with FSDP. }
\end{figure}

Distinct from these works, we explore a promising yet underutilized direction: \textbf{lossless compression}. By leveraging data redundancy and entropy coding (e.g., Huffman~\cite{huffman2007method}, ANS~\cite{duda2015use}), lossless compression provides a bit-exact representation without any information loss. Our insight stems from the statistical properties of LLM training tensors, including activations~\cite{xicoat2025}, gradients~\cite{shi2019understanding,lin2024stochastic}, and parameters~\cite{si2025unveiling}. Driven by the extensive use of Layer Normalization and well-conditioned optimization landscapes, these tensors typically follow Gaussian or log-normal distributions~\cite{gauss1, si2025unveiling, gauss3, qlora, SqueezeLLM} (some experimental results can also be found in \S\ref{subsec:norm}). In the de-facto BFloat16 (BF16) format~\cite{kalamkar2019study}, this statistical regularity manifests as a highly concentrated distribution of exponent values, where a small subset of exponents accounts for the vast majority of the data. To validate this potential, we analyze the communication patterns of a Qwen3-A22B~\cite{qwen3} Mixture-of-Experts (MoE) model, specifically focusing on the bandwidth-intensive All-to-All operations required at MoE layers. As illustrated in Fig.~\ref{fig:comp_ratio}, tracking the communication volume over 40,000 training iterations reveals that the original data volume is $1.33\times$ larger than its losslessly compressed counterpart. Furthermore, this compression gain increases as training progresses, suggesting substantial potential for mitigating communication bottlenecks in large-scale runs.

However, realizing this potential in practice is challenging. The core obstacle is that the computational overhead of compression often outweighs the bandwidth savings. While lossless compression has been applied in checkpoints~\cite{waddington2025lossless,zipnn} or inference~\cite{dfloat11,zipserv}, applying it to the critical path of training communication requires ultra-low latency. General-purpose GPU compressors like DietGPU~\cite{dietgpu} and nvCOMP~\cite{NVIDIA_nvcomp_2025} are ill-suited for this regime. They fail to leverage communication-aware memory hierarchies and incur significant kernel-launch overheads. As shown in Fig.~\ref{fig:compress_compare}, this leads to a ``performance cliff'': despite reducing data volume, the substantial compression and decompression overheads make these methods slower than direct uncompressed communication (NCCL) for FSDP workloads, effectively turning a network bottleneck into a computational one.

% To bridge the gap between theoretical compression ratios and practical system speedup, we present \modelname{}, a high-performance lossless communication library for LLM training. \modelname{} serves as a drop-in replacement for standard collectives, achieving speedups through three key innovations. First, we introduce \textbf{theoretically grounded exponent coding} (\S\ref{sec:top7}). Instead of costly dynamic histogram construction, we exploit the statistical normality of LLM tensors to derive theoretical bounds on exponent concentration. This allows us to compute static, compact lookup tables, eliminating sorting overheads for deterministic, low-latency compression. Second, we implement \textbf{communication-aware kernel design} (\S\ref{sec:kernels}). By fusing compression with data movement and designing data layouts that align with collective patterns, we eliminate redundant transformations. Our implementation meticulously manages global memory access and shared memory banking, pipelining execution to fully mask compression latency. Finally, addressing the load imbalance in MoE training, we propose \textbf{adaptive strategies} (\S\ref{sec:implement}) for All-to-All and a dynamic switcher for Reduce-Scatter that adjusts to runtime system characteristics.

To address the issue, we present \modelname{}, a lossless communication collective library (\S\ref{sec:zipccldesign}) for LLM training, which bridges the gap between the theoretical compression potential of near‑Gaussian LLM tensors and practical GPU performance with three key innovations: \ding{182} We propose a theoretically grounded exponent coding technique (\S\ref{sec:top7}) that exploits the normality of LLM tensors to bypass dynamic histogram construction. By deriving theoretical bounds on exponent concentration in normally distributed floating‑point data, we directly compute a compact lookup table for exponent remapping, eliminating the need for online sorting and enabling deterministic, low‑latency compression. \ding{183} We design GPU-optimized compression and decompression kernels (\S\ref{sec:kernels}) with a communication-aware data layout, which is used to carefully design memory access patterns and pipelines. By leveraging the characteristics of communication operations, we design data layouts and access patterns that eliminate redundant data transformations. In the kernel implementation, we carefully account for global memory access costs and shared memory banking effects, and implement pipelined execution to overlap computation with data movement. \ding{184} We propose adaptive communication strategies for All-to-All operations (\S\ref{sec:implement}) in MoE model training under imbalanced expert computation loads, and we design a dynamic switcher for Reduce-Scatter that adapts to system characteristics.

We implement \modelname{} as a drop‑in replacement for standard NCCL collectives and evaluate it on a 64‑GPU cluster using representative MoE (DeepSeek‑V3, Qwen‑MoE) and dense (Llama3‑8B) models. Experiments show that \modelname{} reduces communication time by up to \(1.35\times\) and achieves end‑to‑end training speedups of \(1.16\)–\(1.18\times\) over state-of-the-art training systems including Megatron-LM~\cite{narayanan2021efficient} and TorchTitan~\cite{liang2025torchtitan}.

\section{Preliminaries and Motivations}
% \begin{figure}[!tb]
% 	\centering
% 		\includegraphics[width=0.48\textwidth]{figures/ep.pdf}
% 	\caption{An illustration of Expert Parallelism (EP). }
% 	\label{fig:ep}
% \end{figure}

% \begin{figure}[!t]
% 	\centering
%  \begin{subfigure}[b]{0.48\textwidth}
% 		\centering
% 		\includegraphics[width=\textwidth]{figures/fsdp_f.pdf}
% 		\caption{Forward}
% 		\label{fig:fsdp_f}
% 	\end{subfigure}
%     	 \begin{subfigure}[b]{0.48\textwidth}
% 		\centering
% 		\includegraphics[width=\textwidth]{figures/fsdp_b.pdf}
% 		\caption{Backward}
% 		\label{fig:fsdp_b}
% 	\end{subfigure}
% 	\caption{The illustration for the forward and backward process of FSDP.}
% 	\label{fig:fsdp}
% \end{figure}

\subsection{Parallelism Paradigms}
Parallelism is an essential technique for large‑scale LLM training. Common parallelism strategies include Data Parallelism (DP)~\cite{DBLP:dean2012large}, Pipeline Parallelism (PP)~\cite{narayanan2021efficient}, Tensor Parallelism (TP)~\cite{DBLP:dean2012large,narayanan2021efficient}, Expert Parallelism (EP)~\cite{DBLP:Shazeer2017outrageously, hwang2023tutel}, and Fully Sharded Data Parallelism (FSDP)~\cite{zhao2023pytorchfsdp}. 
% Since this paper primarily validates our methods on EP and FSDP, we briefly introduce these two approaches below.

\textbf{Expert Parallelism} is specifically designed for Mixture-of-Experts (MoE) models, where different experts are distributed across multiple devices. Each device hosts a subset of experts and processes only the tokens assigned to its local experts. Communication occurs via an All‑to‑All operation to dispatch tokens to targeted devices based on the gating function's decisions and to combine tokens after local experts computation. EP maximizes device utilization in MoE training by allowing multiple experts to be activated simultaneously across the system.

\textbf{Fully Sharded Data Parallelism} is an advanced form of data parallelism that shards model parameters, gradients, and optimizer states across all devices. Unlike classic DP, which replicates the entire model on each GPU, FSDP distributes the model's components, thereby significantly reducing per‑device memory footprint. During the forward pass, each device gathers the necessary parameters via All-Gather for its local computation; in the backward pass, each device also gathers parameters to compute gradients which are then synchronized via Reduce‑Scatter, and optimizer updates are performed on the sharded parameters. FSDP enables training of very large models that would otherwise exceed the memory of a single accelerator.

\textbf{Data Parallelism} synchronizes model states by replicating the full model on each worker. During the backward pass, after local gradients are computed, an All-Reduce operation is triggered across all replicas to aggregate and average the gradients for consistent parameter updates.

\textbf{Tensor Parallelism} partitions individual weight matrices within a layer across multiple devices. To ensure mathematical equivalence to a non-partitioned layer, communication is required in both forward and backward passes. Specifically, a TP implementation invokes an All-Reduce after each block’s linear transformation to synchronize partial results before the next activation function.

\begin{figure}[!t]
	\centering
		\includegraphics[width=0.38\textwidth]{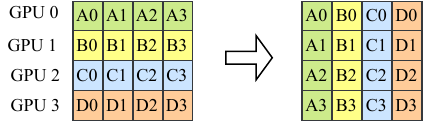}
	\caption{A 4-GPU example of the All-to-All collective. }
	\label{fig:a2a_n}
\end{figure}
\begin{figure}[!t]
	\centering
		\includegraphics[width=0.48\textwidth]{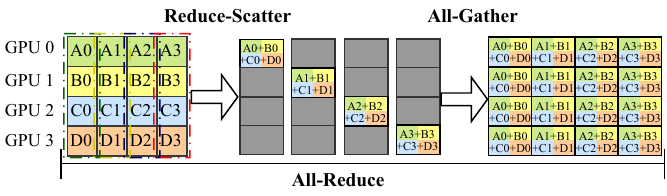}
	\caption{A 4-GPU example of All-Gather, Reduce-Scatter and All-Reduce collectives. }
	\label{fig:ar_n}
\end{figure}
\subsection{Communication Collectives}\label{subsec:collective}
Communication collectives~\cite{nccl} are essential primitives for distributed training of LLMs using parallelism paradigms enabling efficient data exchange and synchronization across multiple devices. 

\textbf{All-to-All}.
The All-to-All collective enables each process in a group to send distinct data to every other process and receive different data from all processes. As shown in Fig.~\ref{fig:a2a_n}, All-to-All divides the data into chunks corresponding to the number of devices, sequentially sends these chunks to each device, and simultaneously receives data from every other device.

\textbf{Reduce‑Scatter}.
Reduce‑Scatter performs an element‑wise reduction (e.g., sum) over a tensor distributed across all processes, then scatters the resulting reduced slices so that each process receives a distinct portion as shown in the right part of Fig.~\ref{fig:ar_n}. It is widely used in TP to aggregate partial results across devices and in FSDP to combine partial weight or gradient across devices.

\textbf{All‑Gather}.
The All‑Gather operation collects distinct data blocks from each process and concatenates them, delivering the complete assembled tensor to every participant as shown in the left part of Fig.~\ref{fig:ar_n}. This collective is commonly employed in TP and FSDP. Each worker initially holds a unique segment; after the collective, all workers possess the full concatenated output.

\textbf{All-Reduce}.
The All-Reduce collective operation is fundamental in distributed deep learning for synchronizing gradients or other tensors across all processes in a group. It combines values from all processes using a reduction operation and distributes the result back to all processes. This operation is crucial for gradient synchronization in data parallel training, ensuring all model replicas maintain consistent parameter updates. All-Reduce can be viewed as a combination of Reduce-Scatter followed by an All-Gather~\cite{rabenseifner2004optimization,thakur2005optimization} as shown in Fig.~\ref{fig:ar_n}. During Reduce-Scatter, each device's data is first divided into chunks, allowing every device to obtain one chunk of the reduced result across devices. The All-Gather then ensures that every card possesses all the reduced data.

\subsection{Floating-Point Format}
In LLM training, the most commonly used floating‑point format is BFloat16 (BF16)~\cite{kalamkar2019study, mix_precision}, which strikes a balance between memory efficiency and numerical robustness. It is widely deployed in production‑scale models, including Llama3~\cite{dubey2024llama}, Qwen~\cite{qwen3}, and Mistral~\cite{jiang2024mixtral}. 

A BF16 value is composed of 1 sign bit, 8 exponent bits, and 7 mantissa bits. Its numeric representation follows:
\[
\text{BF16}(x) = (-1)^{\text{sign}} \times 2^{\text{exponent} - 127} \times (1.\text{mantissa}).
\]
This design retains the full 8‑bit exponent range of IEEE 754 single‑precision (FP32) while reducing the mantissa to 7 bits. Compared to the half‑precision FP16 format (1‑5‑10), BF16 provides a substantially wider dynamic range, which significantly mitigates the risk of overflow and underflow in large‑scale model training.

Consequently, most communication operations during training are performed in BF16. Nevertheless, a subset of operations, such as optimizer updates and gradient accumulation, still relies on higher‑precision FP32 arithmetic to maintain numerical stability and convergence fidelity.

\begin{table}[!t]
\centering
\small
\caption{Percentage of the top $k$ most frequent exponents covering all exponents in BF16 data sampled from a normal distribution.}
\label{tab:exponent_coverage_horizontal}
\begin{tabular}{@{}c *{7}{c} @{}}
\toprule
\textbf{Top $k$} & 1 & 2 & 3 & 4 & 5 & 6 & 7 \\
\midrule
\textbf{Coverage (\%)} & 29.9 & 57.2 & 75.7 & 85.5 & 90.4 & 95.0 & 97.5\\
\bottomrule
\end{tabular}
\end{table}

\subsection{Motivation}
\textbf{Underutilized data distribution in LLM communication compression.}
A large fraction of the data exchanged during LLM training, including activations, gradients, and weight parameters, are approximately Gaussian or log‑normal distributions~\cite{qlora,SqueezeLLM,gauss1,si2025unveiling,gauss3}, which is also observed in\S\ref{subsec:norm}. This statistical regularity is especially pronounced in well‑normalized intermediate tensors and has been observed across different model architectures and training stages. While general compression algorithms ignore such distributional priors, this property can be exploited to accelerate compression. In particular, for exponent‑based formats like BF16, the distribution of exponent values is highly concentrated; the top‑7 most frequent exponents often cover over 97\% of the data as shown in Table~\ref{tab:exponent_coverage_horizontal}. By theoretically bounding the concentration of exponents in normally distributed floating‑point data, we can directly derive the lookup table required for exponent remapping without expensive histogram collection and sorting.

\textit{Thus, a distribution‑aware compression method that leverages the inherent normality of LLM tensors can bypass costly statistical gathering steps and enable near‑optimal, low‑latency compression fully on the GPU.}

\textbf{Inefficient communication‑aware lossless compression for LLM training.} Communication accounts for a substantial portion of the end‑to‑end training time in distributed LLM training. Lossless compression can effectively reduce the volume of transmitted data and thus alleviate communication bottlenecks. However, existing GPU‑accelerated lossless compression schemes often fail to fully exploit GPU architectural and communication characteristics, resulting in suboptimal throughput and high compression latency. As introduced in \S\ref{subsec:collective}, efficient communication collectives require splitting data into chunks before transmission. However, compression and decompression kernels typically reorganize the data layout after processing, preventing direct chunk‑wise splitting. For example, to support All‑to‑All, the data must be manually partitioned and compressed per chunk, which introduces significant overhead. Furthermore, existing compression methods are insufficiently optimized for GPU execution: they lack careful design of data access and storage patterns for global and shared memory, resulting in inefficient memory transactions and limited overlap between computation and data movement.

\textit{Therefore, designing communication-aware lossless compression and decompression GPU kernels that fully align with GPU memory hierarchies, minimize kernel launch overhead, and efficiently overlap compression with memory access is essential to achieve practical end‑to‑end speedups in LLM training.}

\begin{figure}[!t]
	\centering
		\includegraphics[width=0.48\textwidth]{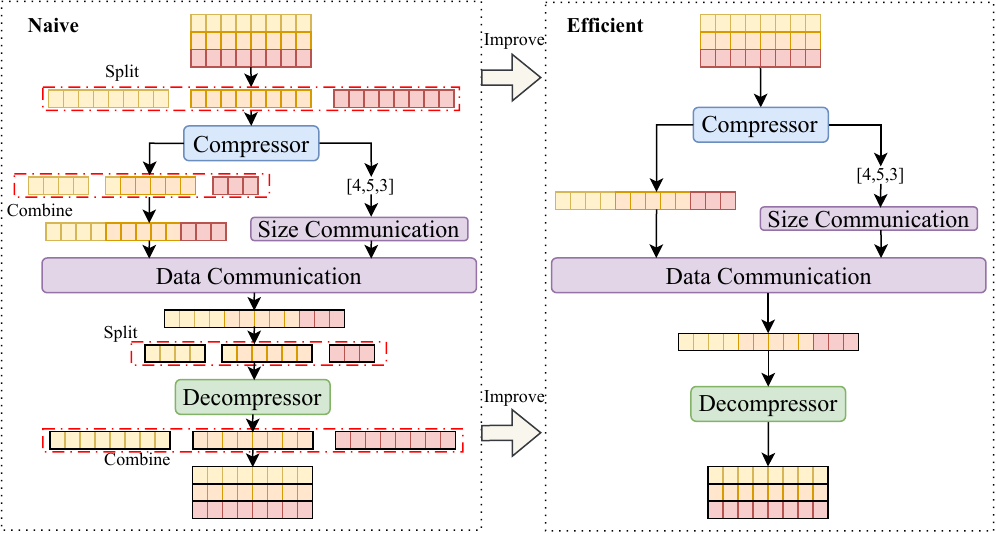}
	\caption{Comparison of the naive implementation of lossless compressed communication collectives (left) and our efficient \modelname{} design (right). }
	\label{fig:zccl}
\end{figure}
\begin{figure}[!t]
	\centering
 \begin{subfigure}[b]{0.48\textwidth}
		\centering
		\includegraphics[width=\textwidth]{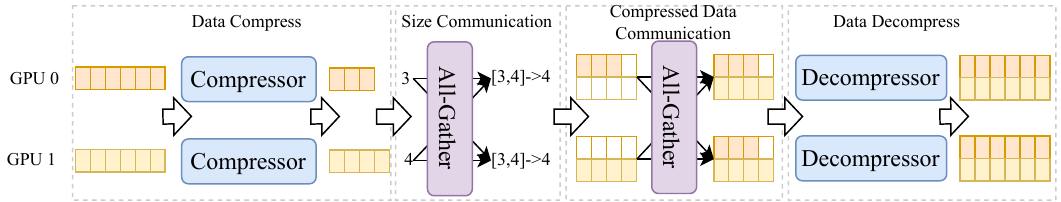}
		\caption{Zipped All-Gather}
		\label{fig:ag}
	\end{subfigure}
    	 \begin{subfigure}[b]{0.48\textwidth}
		\centering
		\includegraphics[width=\textwidth]{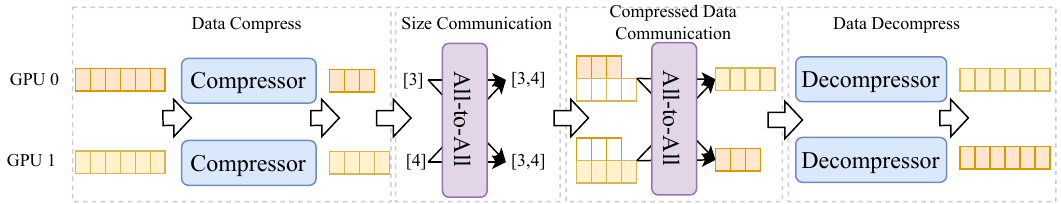}
		\caption{Zipped All-to-All Design 1}
		\label{fig:a2a1}
	\end{subfigure}
    \begin{subfigure}[b]{0.48\textwidth}
		\centering
		\includegraphics[width=\textwidth]{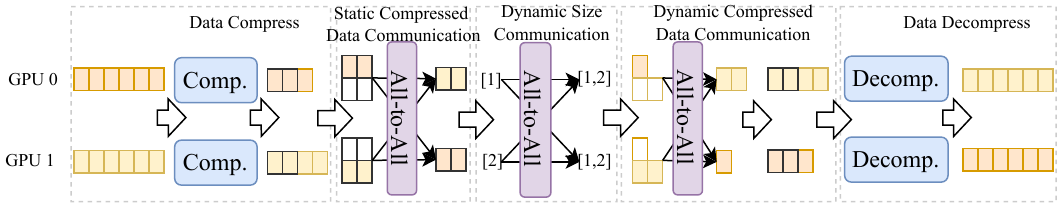}
		\caption{Zipped All-to-All Design 2}
		\label{fig:a2a2}
	\end{subfigure}
      \begin{subfigure}[b]{0.48\textwidth}
		\centering
		\includegraphics[width=\textwidth]{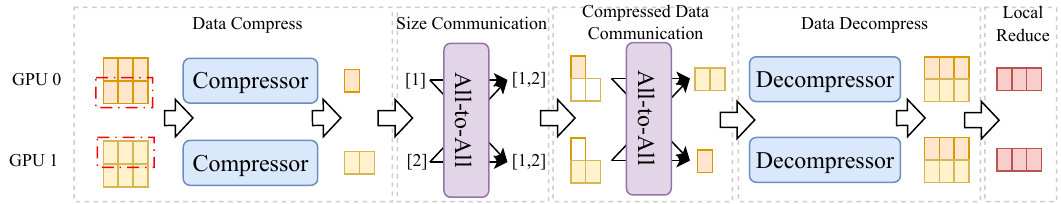}
		\caption{Zipped Reduce-Scatter}
		\label{fig:rs}
	\end{subfigure}
	\caption{The workflow of our \texttt{zipcclAllGather}, \texttt{zipcclAlltoAll}, and \texttt{zipcclReduceScatter}.}
	\label{fig:zccl_full}
\end{figure}

% \section{Zipped Communication Collectives}\label{sec:zipccldesign}
\section{\modelname{}: Library Design}\label{sec:zipccldesign}
% To reduce communication overhead in LLM training, we design a fast lossless compressor and decompressor tailored for data following a normal distribution, and based on it, we propose a family of communication primitives that directly integrate this fast compressor into the communication pipeline.
To seamlessly support the communication collectives for LLM training, we design our \modelname{}, where each zipped collective is one API that is compatible with NCCL. First, to maximally reduce the extra operations required for data compression and decompression, we implement our lossless data compressor and its decompressor together with collectives, as shown in Fig.~\ref{fig:zccl}. Compared to the naive implementation (left in Fig.~\ref{fig:zccl}), which requires two extra split operations and two combine operations to enable compressed data collectives, our \modelname{} design (right in Fig.~\ref{fig:zccl}) eliminates these extra operations using customized GPU kernels (\S\ref{sec:kernels}).
Specifically, data is compressed before transmission, followed by decompression. Because the compressed size depends on data distribution, each collective performs a lightweight synchronization step to exchange only the metadata (i.e., compressed sizes) prior to data transfer (details in the following subsections for each collective). 

We implement \modelname{} as a library of collectives atop NCCL and follow the same input format requirements as NCCL collectives, thereby minimizing extra data movement and kernel overhead that would be needed for format alignment. Since All-Reduce can be viewed as a Reduce-Scatter operation followed by an All-Gather operation, we focus on three collectives: All‑Gather, All‑to‑All, and Reduce-Scatter with our lossless compression; that is: \texttt{zipcclAllGather}, \texttt{zipcclAlltoAll}, and \texttt{zipcclReduceScatter}, respectively.
% that address their unique communication patterns and performance challenges. 

\subsection{\texttt{zipcclAllGather}}

As shown in Fig.~\ref{fig:ag}, zipped All‑Gather enhances the standard All‑Gather with compression and decompression stages. Each worker first applies a fast lossless compressor to its local data. Since the compressed sizes vary between workers, an All-Gather operation should be used to communicate the compressed data size for each worker.
% “All‑Gather of sizes” is performed where each worker exchanges a single integer representing its compressed data size. 
This operation only introduces negligible overhead, as only an integer is all-gathered, but it enables precise buffer allocation and efficient scheduling of the subsequent data movement. The compressed blocks are then gathered using another All‑Gather operation, after which each worker decompresses all received blocks to reconstruct the full tensor. 

\subsection{\texttt{zipcclAlltoAll}}

All‑to‑All is a bandwidth‑heavy collective operation that plays a key role in training MoE models, as it redistributes tokens among the different experts. As shown in Fig.~\ref{fig:a2a1} and Fig.~\ref{fig:a2a2}, we propose two practical designs for \texttt{zipcclAlltoAll}. The first design (Fig.~\ref{fig:a2a1}) follows a four‑phase approach: compress locally, exchange compressed sizes via an additional All‑to‑All, then transmit the compressed data using the size information and decompress the received data. The second design (Fig.~\ref{fig:a2a2}) splits compressed data into a ``static part'' (whose size is known before compression and is unaffected by data distribution) and a ``dynamic part'' (whose compressed size varies with the input). The workflow follows: first, compress locally; second, transmit the static part via a conventional All‑to‑All; third, exchange the sizes of the dynamic parts across workers; fourth, transmit the dynamic data itself using the exchanged size information; finally, decompress all received blocks.  
This ordering is motivated by load‑imbalance patterns commonly observed in MoE training: because expert computation times can vary significantly across devices, workers may become ready for communication at different times. By front‑loading the transmission of the static part, whose size is fixed and known a prior, we allow faster devices to start useful communication earlier while slower devices catch up. A detailed discussion of the load‑imbalance problem and how our design addresses it is provided in \S\ref{sec:imbalance}.

\subsection{\texttt{zipcclReduceScatter}}

Directly integrating compression into a native Reduce‑Scatter is challenging because decompression is required before every reduction step. Instead, we implement \texttt{zipcclReduceScatter} by decomposing the operation into a \texttt{zipcclAlltoAll} followed by a local reduction, as illustrated in Fig.~\ref{fig:rs}. Each worker performs a \texttt{zipcclAlltoAll} operation to send and receive data with other workers. After that, each worker decompresses the data and sums the results locally. This decomposition shifts the reduction from the network to the device, reusing the efficient \texttt{zipcclAlltoAll}. However, the performance of this decomposed form depends on cluster characteristics; on networks with high bisection bandwidth, a native Reduce‑Scatter may be faster. Therefore, we equip the collective with an ``adaptive switcher'' (\S\ref{sec:switch}) that profiles the cluster at initialization and selects between the decomposed path and the native Reduce‑Scatter based on a cost model that considers measured bandwidth, latency, and compression ratio. This ensures robustness across different hardware configurations, always matching or outperforming the baseline while providing substantial gains when compression is effective.

\begin{figure}[!t]
	\centering
		\includegraphics[width=0.48\textwidth]{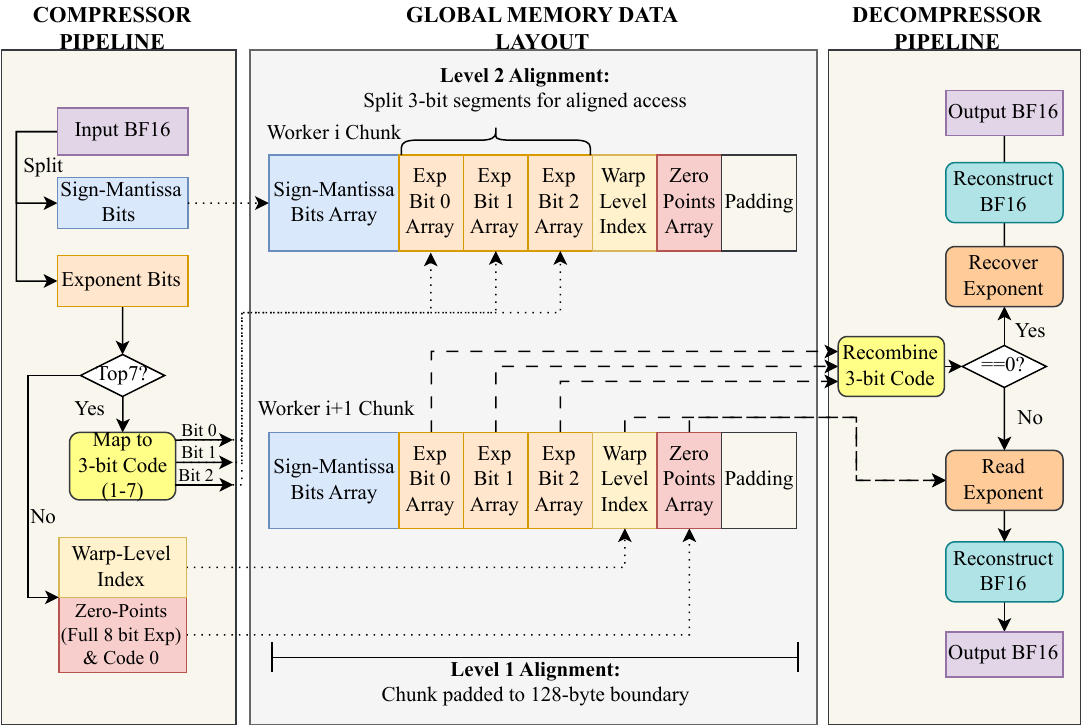}
	\caption{The workflows of our Compressor and Decompressor on the GPU. }
	\label{fig:coderglobal}
\end{figure}
% \begin{figure}[!t]
% 	\centering
%  \begin{subfigure}[b]{0.48\textwidth}
% 		\centering
% 		\includegraphics[width=\textwidth]{figures/compress.pdf}
% 		\caption{Compression}
% 		\label{fig:comp}
% 	\end{subfigure}
%     	 \begin{subfigure}[b]{0.48\textwidth}
% 		\centering
% 		\includegraphics[width=\textwidth]{figures/decompress.pdf}
% 		\caption{Decompression}
% 		\label{fig:decomp}
% 	\end{subfigure}
% 	\caption{The illustration of Compression and Decompression.}
% 	\label{fig:compdecomp}
% \end{figure}
\section{Communication-Aware Lossless Compression and Decompression}\label{sec:compression-kernels}
% Lossless compression can reduce communication volume, thereby lowering communication overhead and improving LLM training efficiency. However, efficiently implementing lossless compression has remained a major challenge. Previous studies have observed significant redundancy in the exponent bits of floating‑point numbers. Taking BF16 as an example — composed of 1 sign bit, 8 exponent bits, and 7 mantissa bits — the exponent bits exhibit substantial redundancy, especially in normally distributed data. Table1 shows the percentage of data covered by the top‑k most frequent exponent values in a normally distributed BF16 dataset: only 7 exponents are needed to cover 97\% of the data. This characteristic enables lossless compression of BF16, but existing approaches require counting and sorting the frequency of each exponent value, which is computationally expensive and difficult to accelerate on GPUs and thus remained implemented on CPU. To address this issue, we propose a theoretically grounded method for rapidly computing the top‑k exponents in normally distributed data. 
We focus on lossless compression targeting the exponent bits of floating‑point numbers, which has been demonstrated by numerous studies~\cite{neuzip,zipserv,dietgpu} to contain significant redundancy. For normally distributed data, the coverage of the top‑k most frequent exponent values is shown in Table~\ref{tab:exponent_coverage_horizontal}: just 7 exponents can cover 97\% of the data. Therefore, we encode exponents using these 7 values plus an additional ``zero‑point'' flag (indicating an exponent outside the top‑7 set), requiring only three bits per exponent. For the BF16 format commonly used in LLM training, this lossless compression reduces the exponent storage from 8 bits to 3 bits, thereby lowering the total bit‑width per value from 16 bits to 11 bits, achieving a theoretical compression ratio of up to 31\%.

As illustrated in Fig.~\ref{fig:coderglobal}, our compression pipeline bifurcates the data stream into decoupled segments. First, each floating‑point number is split into its exponent bits and the remaining bits (sign and mantissa). The sign and mantissa bits are preserved in their raw form to ensure a fixed-stride memory layout, which allows the hardware to perform vectorized and coalesced memory accesses and stores without the overhead of bit-shuffling. Concurrently, for the exponent, we check whether its value belongs to the predetermined set of top‑7 exponents. If yes, it is mapped to a 3‑bit code (001‑111); otherwise, it is assigned the special code 0 (000), and the full 8‑bit exponent is stored separately as a ``zero‑point.'' Third, we aggregate the counts of these zero‑points to generate warp-level indices for efficient storage and to determine the final compressed size for communication. It is worth noting that, apart from the data size of the exponent bits corresponding to zero-points, all other data sizes remain fixed. This aligns with the earlier distinction between static compressed data and dynamic compressed data.
The decompression workflow follows a similar process: first, the 3‑bit exponent code is examined; if it is non‑zero, the original exponent is recovered by looking up the corresponding top‑7 value; if it is zero, the full exponent is read from the stored zero‑point list with the generated index. Then, the exponent is recombined with the saved sign‑mantissa bits to reconstruct the original floating‑point value.

\subsection{Determination of Top‑7 Most Frequent Exponent Values}\label{sec:top7}
We introduce a method for efficiently identifying the seven most frequent exponent values in BF16 data. The method avoids expensive frequency counting or sorting by deriving and leveraging a theoretical relationship between a layer's weight distribution and its optimal exponent coverage.

To characterize the distribution of exponents in BF16 data, we seek to identify the optimal base exponent that maximizes the probability coverage within a specific range. By determining the relationship between the weight standard deviation $\sigma$ and this optimal exponent $x$, we can quantify the expected redundancy and concentration in the exponent field. Although $x$ is discrete in practice, we treat it as a continuous variable to find the optimal coverage through differentiation.

To analyze the exponent distribution, we first define $x$ as the unbiased exponent value of a BF16 number. Given the BF16 format $v = (-1)^S \times 2^{E - 127} \times (1.m_1...m_7)_2$, the value $x = E - 127$ determines the magnitude range such that $|v| \in [2^x, 2^{x+1})$. Under the assumption that weights follow a Gaussian distribution $w \sim \mathcal{N}(0, \sigma^2)$, we define $P_{\text{win}}(x)$ as the probability that a randomly sampled weight falls within a contiguous window of 8 exponents starting from $x$, covering the magnitude range $[2^x, 2^{x+7})$. By integrating the probability density function over both positive and negative intervals, we obtain:
\begin{equation}
P_{\text{win}}(x) = 2 \int_{2^x}^{2^{x+7}} \frac{1}{\sqrt{2\pi\sigma^2}} e^{-\frac{t^2}{2\sigma^2}} dt.
\end{equation}
This can be simplified using the error function as 
\begin{equation}
P_{\text{win}}(x) = \text{erf}\left(\frac{2^{x+7}}{\sigma\sqrt{2}}\right) - \text{erf}\left(\frac{2^x}{\sigma\sqrt{2}}\right).
\end{equation}
For algebraic simplicity, we define a substituted variable $u(x) = \frac{2^x}{\sigma\sqrt{2}}$. As $2^{x+7} = 128 \cdot 2^x$, the expression becomes $P_{\text{win}}(x) = \text{erf}(128u) - \text{erf}(u)$. To find the base exponent $x$ that maximizes this coverage, we compute the derivative with respect to $x$ using the chain rule, that is
% $\frac{dP}{dx} = \frac{dP}{du} \cdot \frac{du}{dx}$:
\begin{equation}
\frac{dP}{dx} = \frac{dP}{du} \frac{du}{dx}=\left( \frac{2}{\sqrt{\pi}} e^{-(128u)^2} \cdot 128 - \frac{2}{\sqrt{\pi}} e^{-u^2} \right) \cdot (u \ln 2).
\end{equation}
Setting $\frac{dP}{dx} = 0$ for non-zero weights, we obtain the optimality condition $128 e^{-(128u)^2} = e^{-u^2}$. Taking the natural logarithm of both sides yields $\ln 128 - 128^2 u^2 = -u^2$, which simplifies to $(128^2 - 1)u^2 = \ln 128$. Solving for $u$, we find:
\begin{equation}
u = \sqrt{\frac{7 \ln 2}{16383}} \approx 0.0172.
\end{equation}
Substituting the definition of $u$ back into the equation, we derive the relationship $\frac{2^x}{\sigma\sqrt{2}} = \sqrt{\frac{7 \ln 2}{16383}}$. By solving for $x$, we establish the theoretical optimal base exponent as a function of the layer's weight scale $\sigma$:
\begin{equation}
x_{\text{opt}} = \log_2(\sigma) + \frac{1}{2} \log_2 \left( \frac{14 \ln 2}{16383} \right) \approx \log_2(\sigma) - 5.35.
\end{equation}
This result demonstrates that the exponent window providing maximum coverage is centered around a value determined primarily by $\log_2(\sigma)$. In practice, we can utilize high-performance computing via \texttt{torch.std} to efficiently calculate $\sigma$.
% for each layer, thereby determining the optimal exponent distribution peak and enabling effective weight compression.

\subsection{Kernel Design}\label{sec:kernels}
% \subsubsection{Kernel workflow}
% FigureX illustrates the workflow of our compression and decompression kernels. 
% Our compression and decompression workflows are as follows. 
To support the compressor and decompressor in efficiently executing on GPUs, we design dedicated CUDA kernels.

\textbf{Compressor Kernel}:
\ding{182} Global memory to shared memory transfer: Threads load the input BF16 tile from global memory into shared memory using vectorized loads (e.g., "LDG.128").
\ding{183} Shared memory to register transfer and local compression: The tile is moved from shared memory to registers. In registers, each thread separates the sign‑mantissa bits from the exponent bits and performs compression. The partially compressed data is then written back to shared memory.  
\ding{184} Zero‑point counting and warp‑level index: The warp‑level zero‑point index is obtained by first computing a warp‑level prefix sum, then performing inter‑warp synchronization and a block‑level atomic addition to determine each warp’s global index.
\ding{185} Write‑back: The compressed sign‑mantissa array, the three separated 1‑bit exponent arrays, and the warp‑level prefix‑sum metadata are written back to global memory using vectorized stores.

\textbf{Decompressor Kernel}:
\ding{182} Global memory to shared memory transfer of compressed data: The kernel loads the compressed sign‑mantissa array and the three 1‑bit exponent arrays from global memory into shared memory, again using aligned vector loads.  
\ding{183} Shared memory to register transfer and exponent reconstruction: Each warp moves its assigned portion of the sign‑mantissa and exponent bits from shared memory into registers. In registers, each thread performs decompression to reconstruct the original BF16 values and writes them back to shared memory (the warp‑level index and zero-points are directly loaded from global memory due to their small size). 
\ding{184} Shared memory to global memory: The reconstructed BF16 values are then written back to global memory via vectorized stores.
% The warp‑level index, due to its small size, is directly loaded from global memory to the register at first.

\subsubsection{Global Memory Data Layout}
We focus on addressing unaligned memory access patterns in global memory through two alignment strategies.
% In CUDA kernel design, we carefully organize data layouts in both global and shared memory. As shown in Fig.~\ref{fig:coderglobal}, 

In collective operations (e.g., All-to-All), varied compression ratios often shift the boundaries of worker-specific data chunks to arbitrary byte offsets. Such misalignment prevents the hardware from issuing coalesced 128-byte transactions, forcing the memory controller to split single logical requests into multiple physical bus cycles. We eliminate this overhead by padding each worker’s data chunk to a 128-byte boundary, as shown in Fig.~\ref{fig:coderglobal}. This ensures that every worker starts its memory sweep at a DRAM-line-aligned address, enabling the use of \texttt{LDG.E.128} (vectorized float4 loads) to saturate the global memory bandwidth.

We also address the storage of the 3‑bit exponent field used in our compression scheme. Storing these 3-bit values contiguously with the sign and mantissa bits would often lead to unaligned memory accesses, severely degrading efficiency. Therefore, we separate the 3‑bit exponent into three individual 1‑bit segments and store each segment independently, as shown in Fig.~\ref{fig:coderglobal}. When the number of input floating‑point values is a power of two, the start addresses of these three segments also become well‑aligned, enabling coalesced memory operations.
\begin{figure}[!t]
	\centering
		\includegraphics[width=0.48\textwidth]{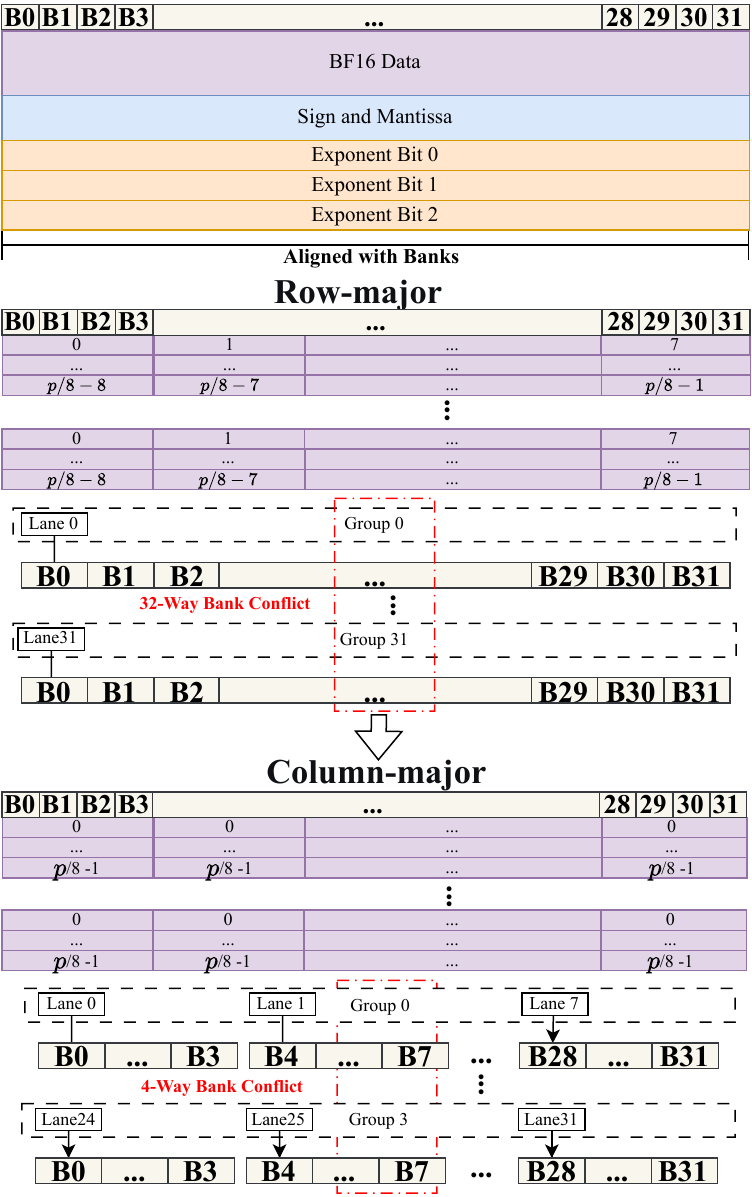}
	\caption{The data layout in shared memory. ``B0, B1, ..., 30, 31'' means ``Bank0, Bank1, ..., Bank30, Bank31''.}
	\label{fig:share}
\end{figure}

\subsubsection{Shared Memory Data Layout}

Shared memory in NVIDIA GPUs is divided into 32 banks, each 4 bytes wide. Concurrent accesses by the 32 threads of a warp to different banks are fully parallelized, while accesses to the same bank cause bank conflicts and serialization. As shown in Fig.~\ref{fig:share}, the data stored in shared memory includes the original floating‑point values in BF16 format, the combined 8‑bit sign‑mantissa fields after splitting, and the compressed 3‑bit exponent values (stored as three separate 1‑bit segments). We first place the original floating‑point array, followed by the 8‑bit sign‑mantissa array, and finally the three 1‑bit exponent segments. Each section is arranged to start at bank~0 and end at bank~31. 

To ensure that global data can utilize `LDG.E.128' (vectorized float4 loads) to saturate global memory bandwidth, we store data in units of a Float4, which corresponds to four bank cells. Fig.~\ref{fig:share} (middle and bottom) also illustrates two data layout schemes for BF16 data, where eight BF16 values constitute one Float4. We set the number of BF16 values processed per thread to $p$. Since we need to store warp‑level zero‑point index data, $p$ cannot be too small; otherwise, the indexing overhead would become excessive. Consequently, a row‑major layout would cause each lane in a warp to access data across all 32 banks, resulting in 32‑way bank conflicts (Fig.~\ref{fig:share} middle). To reduce conflicts, we design a column‑major data layout, which decreases the conflicts to 4‑way (Fig.~\ref{fig:share} bottom).

\subsubsection{Zero‑point Index}

During compression, we count the number of zero points per warp and compute a warp‑level prefix sum of these counts. This prefix sum is stored and used during decompression to quickly locate the start of each warp’s zero‑point exponent data within the global zero‑point array. The computation proceeds hierarchically: first, each thread computes its local count; then warp‑level primitives (e.g., \texttt{\_\_shfl\_up\_sync}) are used to compute the thread‑level prefix sum in each warp; finally, block‑level and global‑level reductions propagate the sums across warps and blocks, yielding a global number of zero points. Moreover, the per-thread prefix sum within each warp is then used to construct the warp-level prefix sum, which is stored and later leveraged to efficiently determine the position of every zero-point’s exponent in the zero-points exponent set.

\subsubsection{Fine‑grained Pipeline}

% We pipeline data movement and compression/decompression operations at two levels. The first level pipelines transfers from global memory to shared memory, while the second level pipelines transfers from shared memory to registers and the actual compression/decompression arithmetic. This is implemented using double‑buffering in both shared memory and registers: while one buffer is being filled from the previous level, the other buffer is being processed by the compression or decompression logic. The overlap effectively hides memory latency and keeps the computational units fully utilized.
To maximize throughput on the NVIDIA GPUs (e.g., Ampere architecture), we implement a two-level hierarchical pipelining strategy that explicitly leverages hardware-level asynchronous primitives. In the first stage, we utilize the Asynchronous Copy Engine via \texttt{cp.async} instructions to facilitate direct data movement from global memory to shared memory. Unlike traditional LDG/STS patterns, this approach bypasses the Register File (RF), significantly reducing register pressure and preventing pipeline stalls caused by register-to-memory dependencies. In the second stage, we decouple the Load/Store Units (LSU) from the Arithmetic Logic Unit (ALU) by employing register-level double-buffering. This allows the LSU to prefetch data into a secondary register bank while the ALU operates on the primary bank. By interleaving these two levels of double-buffering, we effectively hide both memory and instruction latencies, maintaining high functional unit utilization even in the absence of a hardware Tensor Memory Accelerator (TMA).

% \section{Implementation}\label{sec:implement}
\section{Adaptive Communication}\label{sec:implement}
\begin{figure}[!t]
	\centering
		\includegraphics[width=0.48\textwidth]{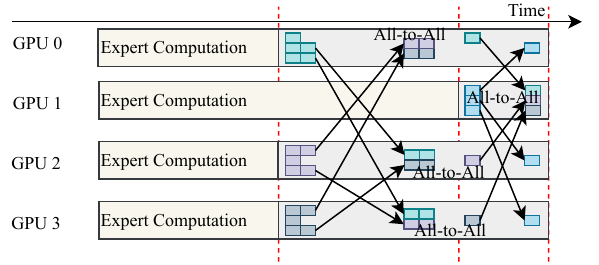}
	\caption{An example of the imbalanced computation workloads in an MoE layer, where GPU 1 requires longer computation time.}
	\label{fig:imbalance}
\end{figure}
\subsection{Imbalanced Workload in MoE Models}\label{sec:imbalance}
In MoE model training, expert load imbalance frequently occurs, leading to significant variations in computation speed across different workers. Some workers complete expert computations quickly, while others are much slower, resulting in uneven communication start times, as shown in Fig.~\ref{fig:imbalance}. This imbalance may be partially mitigated by the subsequent All‑to‑All operation, because All‑to‑All inherently requires all workers to both send and receive data from each other. Workers that start communication earlier can begin exchanging data among themselves, while slower workers initiate their transfers later. 

There are two scenarios that can compensate for computational imbalance. The first arises when communication among workers fails to achieve real peer‑to‑peer concurrency. Consider a simplified scenario where four GPUs need to send and receive $x$ bytes to/from every other GPUs. Suppose that the first GPU starts its communication significantly later than the other three, allowing the remaining three GPUs enough time to fully exchange $x$ bytes among themselves before the first GPU begins. Consequently, when the first GPU finally initiates communication, it only needs to exchange $x$ bytes with each of the other three GPUs. Because true peer‑to‑peer concurrency cannot be achieved, the first GPU may finish its data transfers in less time than a complete four‑GPU All-to-All would take, since the inter‑GPU communication among the remaining three cards no longer impedes the transfers involving the first GPU.

The second scenario arises when early‑starting workers have heavier communication tasks, requiring them to send or receive more data than late‑starting workers. As shown in Fig.~\ref{fig:imbalance}, GPU~1 needs to send and receive $x$ bytes to/from each of the other GPUs, while the other GPUs must exchange $2x$ bytes among themselves. Assuming communication between GPUs is fully P2P‑parallel, the overall communication overhead is determined by the GPU that sends or receives the largest volume of data. In the illustrated example, GPU~1 is the last to become ready. If the other GPUs start communicating earlier, GPU~1 only needs to transfer $3x$ bytes once it begins. However, if all GPUs wait until GPU~1 is ready, then GPUs~0, 2, and 3 each must send and receive $5x$ bytes. Clearly, allowing the other GPUs to start earlier can effectively overlap with the imbalance in expert computation, thereby improving overall throughput.

However, if we adopt the first zipped All‑to‑All design (Fig.~\ref{fig:a2a1}), the initial communication step only exchanges size information, which involves very small data volumes and is insufficient to compensate for the load imbalance caused by uneven expert execution. Therefore, for MoE training, we employ the second scheme (Fig.~\ref{fig:a2a2}), which splits the communication data into static and dynamic parts. The static‑data All‑to‑All helps to absorb the start‑time variation across workers due to expert imbalance.

\subsection{Adaptive Switcher for Reduce-Scatter}\label{sec:switch}
% In large language model training, BF16 precision has become mainstream. 
Reduce‑Scatter is typically used for aggregating activations in TP and gradients in DP and FSDP. Particularly for gradient aggregation, FP32 is often employed to avoid numerical overflow issues such as ``large numbers swallow small numbers.'' However, since our Zipped All‑to‑All + local reduce (\texttt{zipcclReduceScatter}) approach performs no arithmetic during communication—all reduction occurs locally in high precision. It avoids overflow and can safely use BF16 for communication. This reduces communication volume compared to an FP32 Reduce‑Scatter. It is worth noting that when training with FSDP and BF16 precision, the Reduce‑Scatter operation is performed immediately after the gradients of each layer are computed in BF16 format. Although the Reduce‑Scatter itself is executed in FP32, because the initial gradients are in BF16, this process is mathematically equivalent to first performing All‑to‑All communication in BF16 followed by a local reduction in FP32.

Additionally, whether Zipped All‑to‑All is faster than a native Reduce‑Scatter depends on cluster characteristics. To make an adaptive choice, we introduce an adaptive switcher. Following previous works (e.g., FSMoE~\cite{pan2025fsmoe} and PipeMoE~\cite{shi2023pipemoe}), we model the communication costs of Reduce-Scatter (RS) and All-to-All (A2A) with linear functions:
\begin{equation}
    T_{\text{RS}}(d) = \alpha_{\text{RS}} + \beta_{\text{RS}} \cdot d,
\end{equation}
\begin{equation}
T_{\text{A2A}}(d) = \alpha_{\text{A2A}} + \beta_{\text{A2A}} \cdot e \cdot s \cdot d,
\end{equation}
where \(d\) is the data size, \(\alpha\) is the startup latency, \(\beta\) is the per‑byte transmission time, \(e\) is the compression factor of \texttt{zipcclAlltoAll}, and \(s\) is the precision scale factor (e.g., 0.5 if Reduce‑Scatter uses FP16 while \texttt{zipcclAlltoAll} uses BF16). The switcher selects \texttt{zipcclReduceScatter} when \(T_{\text{A2A}}(d) < T_{\text{RS}}(d)\); otherwise, it falls back to the original Reduce-Scatter.

\section{Evaluation}\label{sec:evaluation}
% \begin{table}[]
% 	\centering

%  		\caption{The server configurations in our testbed.}
% 		\label{tab:server-config}
% \resizebox{\linewidth}{!}{%
% \begin{tabular}{ll}
% \hline
% \multicolumn{1}{l}{\textbf{Name}}    & \multicolumn{1}{l}{\textbf{Configuration}} \\ \hline
% \multicolumn{1}{l}{CPU}     & \multicolumn{1}{l}{Dual Intel(R) Xeon(R) Platinum 8358 @ 2.60GHz} \\
% \multicolumn{1}{l}{GPU}     & \multicolumn{1}{l}{8x Nvidia RTX A6000-48G @1.46GHz} \\
% \multicolumn{1}{l}{Memory}  & \multicolumn{1}{l}{512GB DDR4} \\ 
% \multicolumn{1}{l}{NVlink}  & \multicolumn{1}{l}{112.5GB/s (4x)} \\ 
% \multicolumn{1}{l}{PCIe}    & \multicolumn{1}{l}{4.0 (x16)} \\ 
% \multicolumn{1}{l}{Network} & \multicolumn{1}{l}{Mellanox MT28908 @ 200Gb/s} \\ \hline
% \end{tabular}
% % \vspace{-5mm}
% }
% \end{table}
\subsection{Experimental Settings}
% In this section, we describe our experimental testbeds, the baseline training systems, and the LLM models used in our evaluation.

\textbf{Testbeds.}
Experiments are mainly carried out on a 64-GPU cluster comprising four interconnected nodes with 200Gb/s network, each of which is equipped with eight Nvidia RTX A6000 GPUs. We also test the end to end performance on a 16-H800 clusters with 400Gb/s network.

% The details of the server configuration are shown in Table~\ref{tab:server-config}. The main software environments are Ubuntu-22.04, CUDA-12.8, and PyTorch-2.9.1.

\textbf{Baselines. } We integrate our \modelname{} on top of the widely used Megatron-LM~\cite{narayanan2021efficient} and TorchTitan~\cite{liang2025torchtitan} LLM training systems. Megatron-LM supports various MoE architectures that require intensive All-to-All communication, while TorchTitan supports FSDP, which relies heavily on All-Gather and Reduce-Scatter communications. Both training systems utilize PyTorch's built-in NCCL~\cite{nccl} for communication.

\textbf{Real-World LLM Models.} 
We conduct experiments with MoE models in Megatron-LM to evaluate \texttt{zipcclAlltoAll}, and with dense models in TorchTitan using FSDP to evaluate \texttt{zipcclReduceScatter} and \texttt{zipcclAllGather}.
To measure end-to-end training performance on real MoE workloads, we adopt two widely used MoE configurations derived from DeepSeek-V3~\cite{liu2024deepseek} and Qwen3-235B-A22B~\cite{qwen3} (Qwen3-MoE). Given the GPU memory limitations of our platform, we set the number of experts in DeepSeek-V3 to 64 and the total number of layers to 4, while Qwen3-MoE is configured with 16 layers. For dense models, we use Llama3-8B~\cite{dubey2024llama} on TorchTitan with FSDP. The micro-batch size is 1, and the sequence length is 1024.

\begin{figure}[!t]
	\centering
		\includegraphics[width=0.48\textwidth]{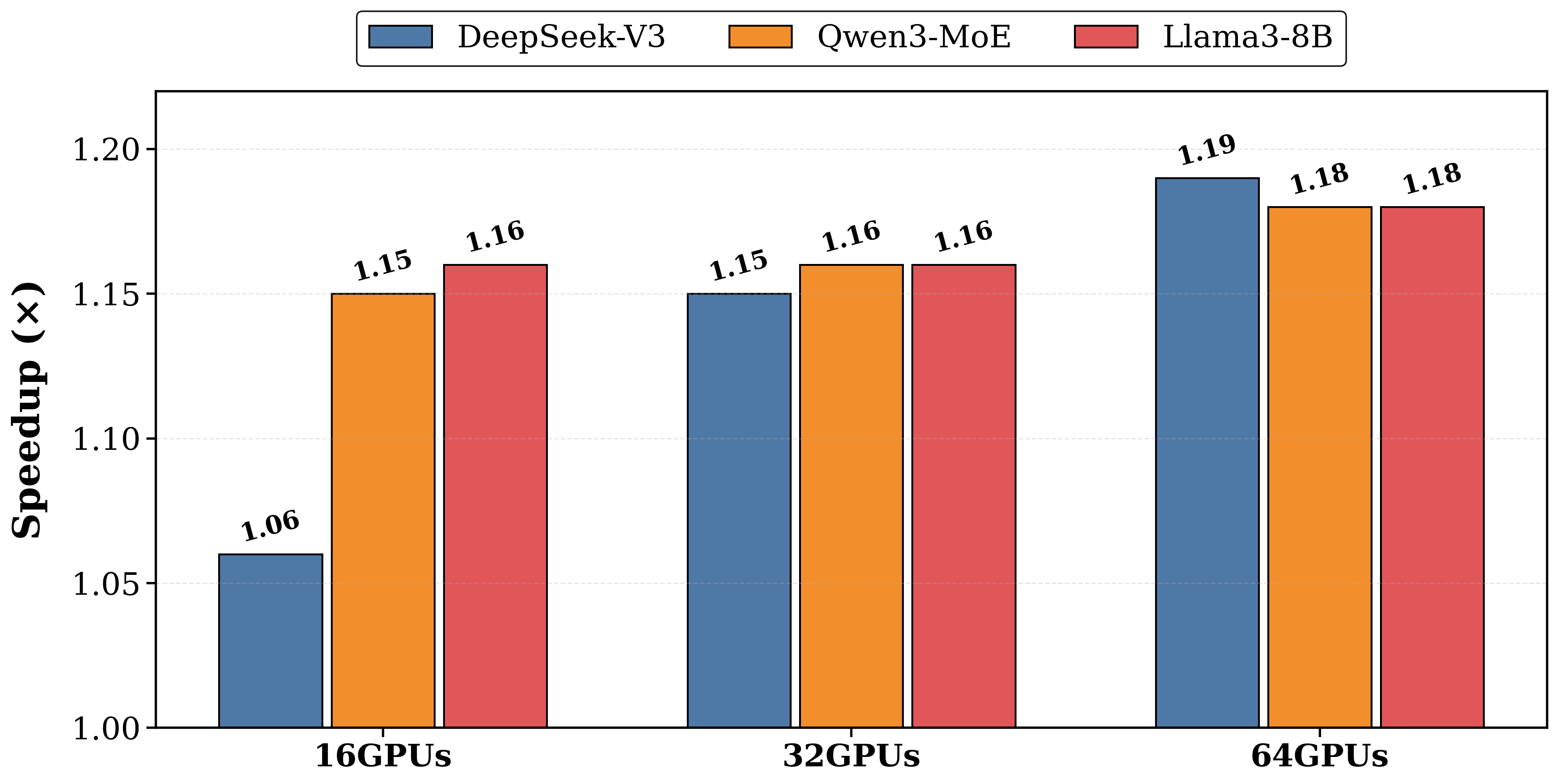}
	\caption{End-to-end speedups of our \modelname{} over baselines on DeepSeek-V3, Qwen3-MoE, and Llama3-8B across different GPU counts. DeepSeek-V3 and Qwen3-MoE are tested on Megatron-LM, while Llama3-8B is evaluated on TorchTitan.}
	\label{fig:e2e}
\end{figure}
\begin{figure}[!t]
	\centering
		\includegraphics[width=0.48\textwidth]{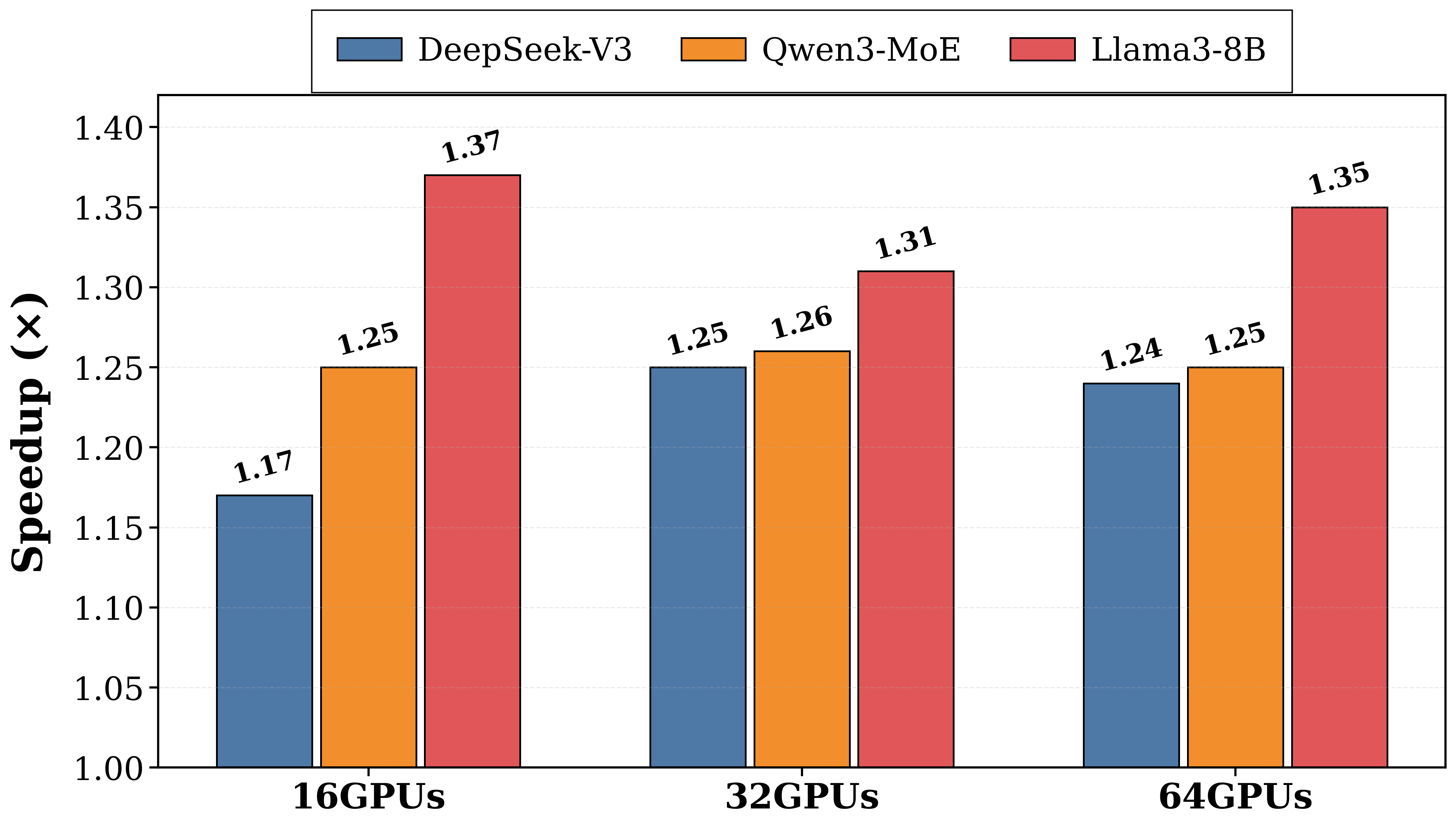}
	\caption{Communication speedups of our \modelname{} over baselines with NCCL on DeepSeek-V3, Qwen3-MoE, and Llama3-8B across different GPU counts. All‑to‑All is measured for DeepSeek‑V3 and Qwen3-MoE, while All‑Gather is measured for Llama3-8B.}
	\label{fig:comm}
\end{figure}
\subsection{End-to-end Training Time Comparison}\label{sec:e2e_compare}
To validate the effectiveness of our method, we compare our \modelname{} on both Megatron‑LM and TorchTitan with NCCL. For Megatron‑LM, we conducted experiments on two MoE models: DeepSeek‑V3 and Qwen3-MoE. For TorchTitan, we tested Llama3-8B. In MoE models, we used DP+EP parallelism, where both DP and EP degrees are equal to the number of GPUs. For Llama3-8B, we employed FSDP, sharding model weights, gradients, and optimizer states across all GPUs. 

The results, as shown in Fig.~\ref{fig:e2e}, demonstrate that our \modelname{} achieves speedups of \(1.16\times\) on DeepSeek‑V3, \(1.16\times\) on Qwen3-MoE, and \(1.18\times\) on Llama3-8B. We also evaluated scaling with GPU counts of 16, 32, and 64. As the number of GPUs increases, the speedups on DeepSeek‑V3 rise from \(1.06\times\) to \(1.16\times\), while Qwen3-MoE remains stable around \(1.16\times\), and Llama3-8B consistently stays near \(1.18\times\).

 We also test the performance on Qwen3-MoE and Llama3-8B on a 16-H800 clusters with 400Gb/s network. Results show that we outperform Megatron-LM by 1.13$\times$ on Qwen3-MoE and outperform TorchTitan by 1.13$\times$ on Llama3-8B. 

\subsection{Communication Time Comparison}
We also separately measured the improvement in communication performance. For DeepSeek-V3 and Qwen3-MoE, we measured the average per‑iteration All‑to‑All communication time. For Llama3-8B, we measured the All‑Gather time, while for Reduce‑Scatter, based on our analysis in \S\ref{sec:switch}, the original Reduce‑Scatter implementation was found to be faster on our platform. The results, as shown in Fig.~\ref{fig:comm}, indicate that our \modelname{} achieves communication speedups of \(1.25\times\) on DeepSeek‑V3, \(1.24\times\) on Qwen3-MoE, and \(1.35\times\) on Llama3-8B over NCCL. We further evaluated scaling across 16, 32, and 64 GPUs. Consistent with the end‑to‑end results, DeepSeek‑V3 shows improvements from \(1.17\times\) to \(1.25\times\) as GPU count increases, Qwen3-MoE remains stable around \(1.25\times\), and Llama3-8B maintains a speedup of approximately \(1.34\times\).
\begin{figure}[!t]
	\centering
		\includegraphics[width=0.48\textwidth]{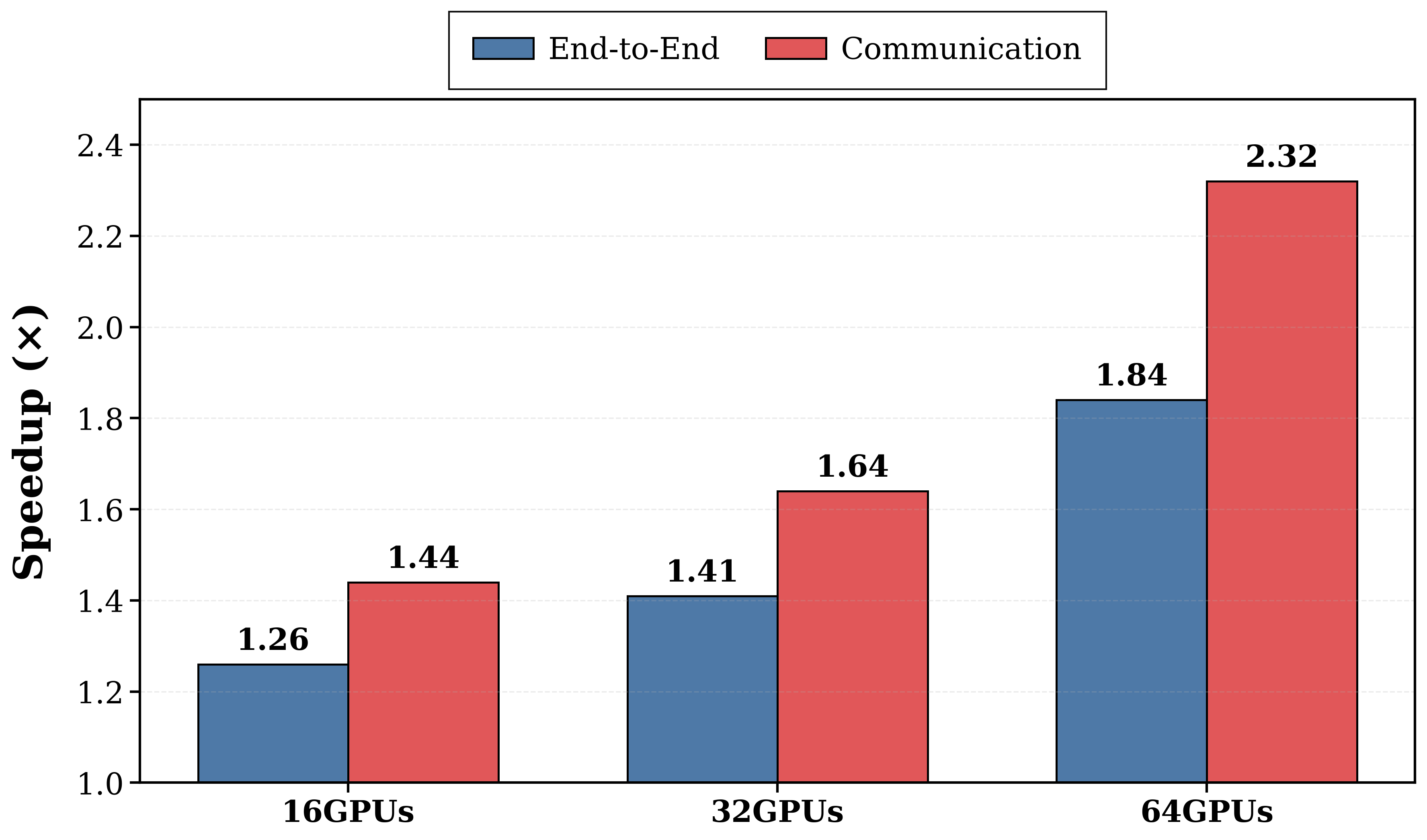}
	\caption{End-to-end and All-to-All communication speedups of our \modelname{} over DietGPU on Qwen3-MoE.}
	\label{fig:DIET}
\end{figure}

\begin{figure}[!t]
	\centering
		\includegraphics[width=0.48\textwidth]{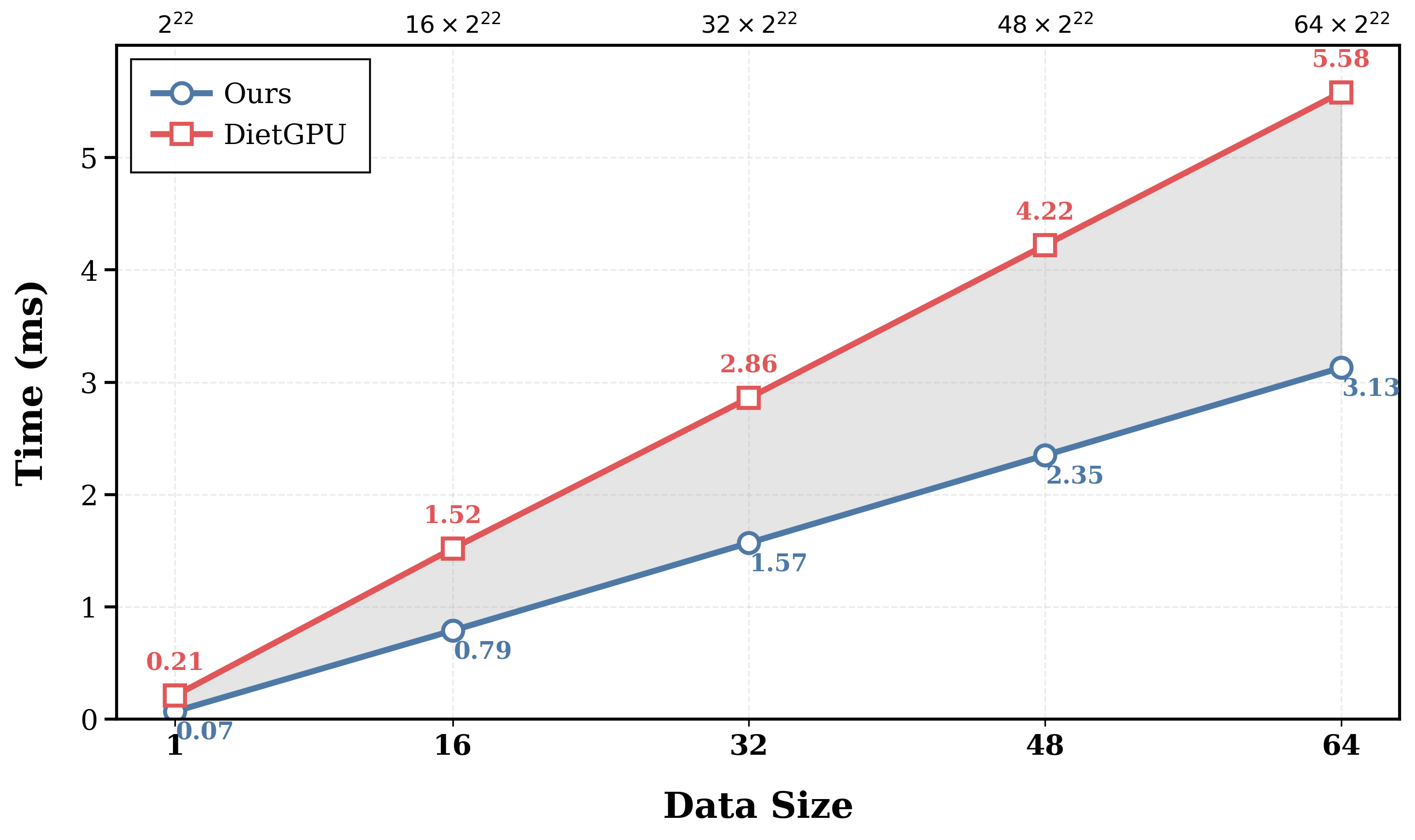}
	\caption{Computation time (compression and decompression) of \modelname{} and DietGPU with different data sizes.}
	\label{fig:zip}
\end{figure}

\subsection{Comparison with DietGPU}
We further compare our method with DietGPU\footnote{\url{https://github.com/facebookresearch/dietgpu}}, a popular GPU‑based lossless compression library. We do not compare with nvCOMP as the data fails to be decompressed when transmitted to other workers. This is likely because nvCOMP generates hidden metadata during compression that is not accessible. Since nvCOMP is closed‑source and its internal metadata format is undocumented, we cannot transmit that information to other workers, making cross‑device decompression infeasible. Additionally, nvCOMP currently lacks support for the BF16 format.

Experiments are conducted on Qwen3-MoE using the same configuration as in \S\ref{sec:e2e_compare}. As shown in Fig.~\ref{fig:DIET}, DietGPU actually increases end‑to‑end training time due to inefficient memory management and excessive data movement overhead, which are not optimized for communication‑centric workflows. In contrast, our method effectively reduces communication volume and accelerates training, achieving 1.26$\times$ to 1.83$\times$ higher end‑to‑end throughput compared to DietGPU.

We also evaluate the pure compression speed, disregarding any data movement and other overheads, of DietGPU versus our compression kernels (\S\ref{sec:compression-kernels}). As illustrated in Fig.~\ref{fig:zip}, our method significantly outperforms DietGPU across both small and large data sizes. Furthermore, we compare the per‑layer All‑Gather time with compression and decompression time required for Llama3-8B training on TorchTitan (due to library conflicts, DietGPU could not be directly integrated into TorchTitan; here we measure the All‑Gather time separately). The results indicate that DietGPU‑compressed All‑Gather performs slower than the baseline NCCL All‑Gather, whereas our approach achieves a 1.35$\times$ speedup.

\begin{figure}[!t]
	\centering
 \begin{subfigure}[b]{0.23\textwidth}
		\centering
		\includegraphics[width=\textwidth]{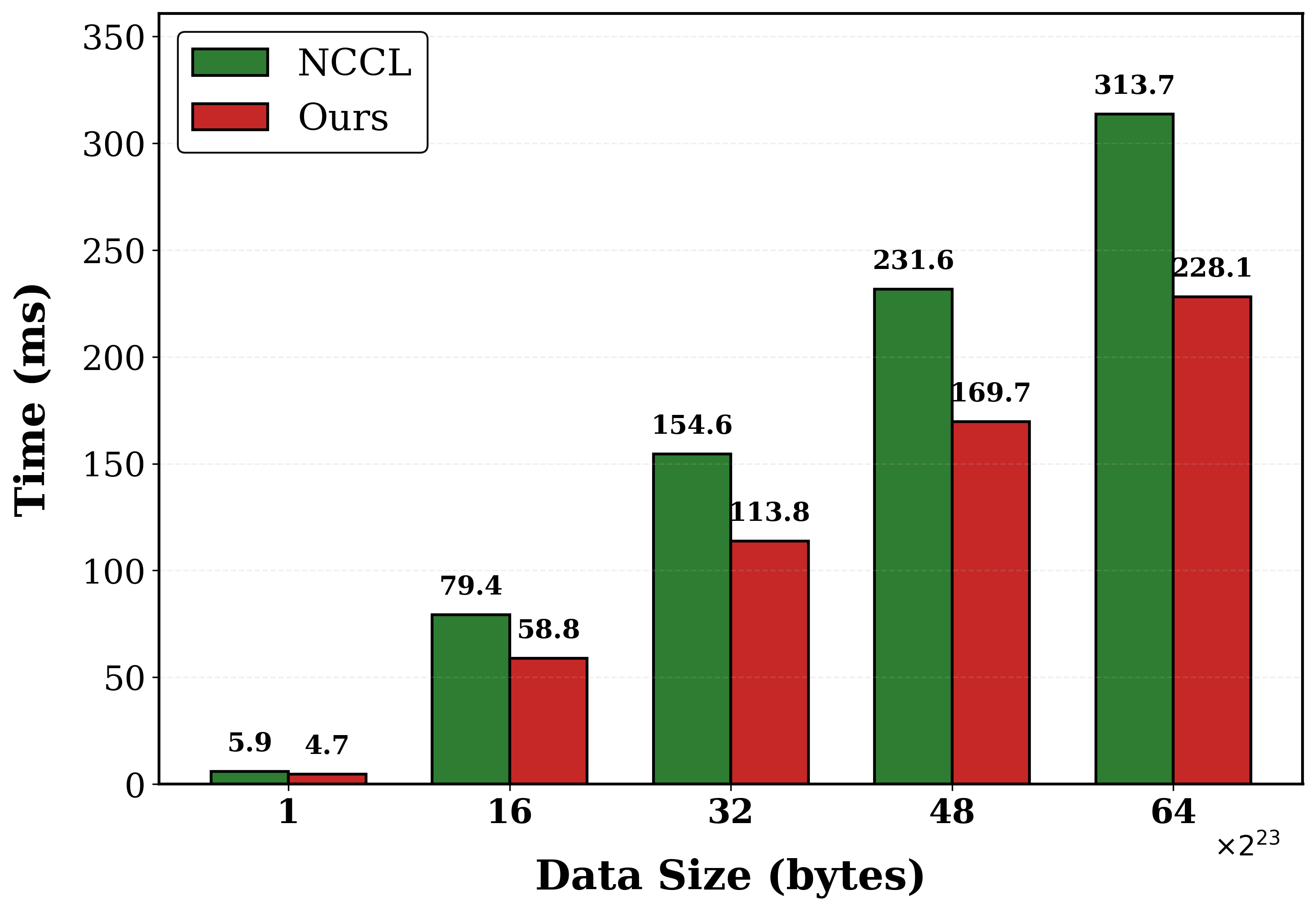}
		\caption{All-to-All time with varied communication volume}
		\label{fig:a2a_vol}
	\end{subfigure}
     \begin{subfigure}[b]{0.23\textwidth}
		\centering
		\includegraphics[width=\textwidth]{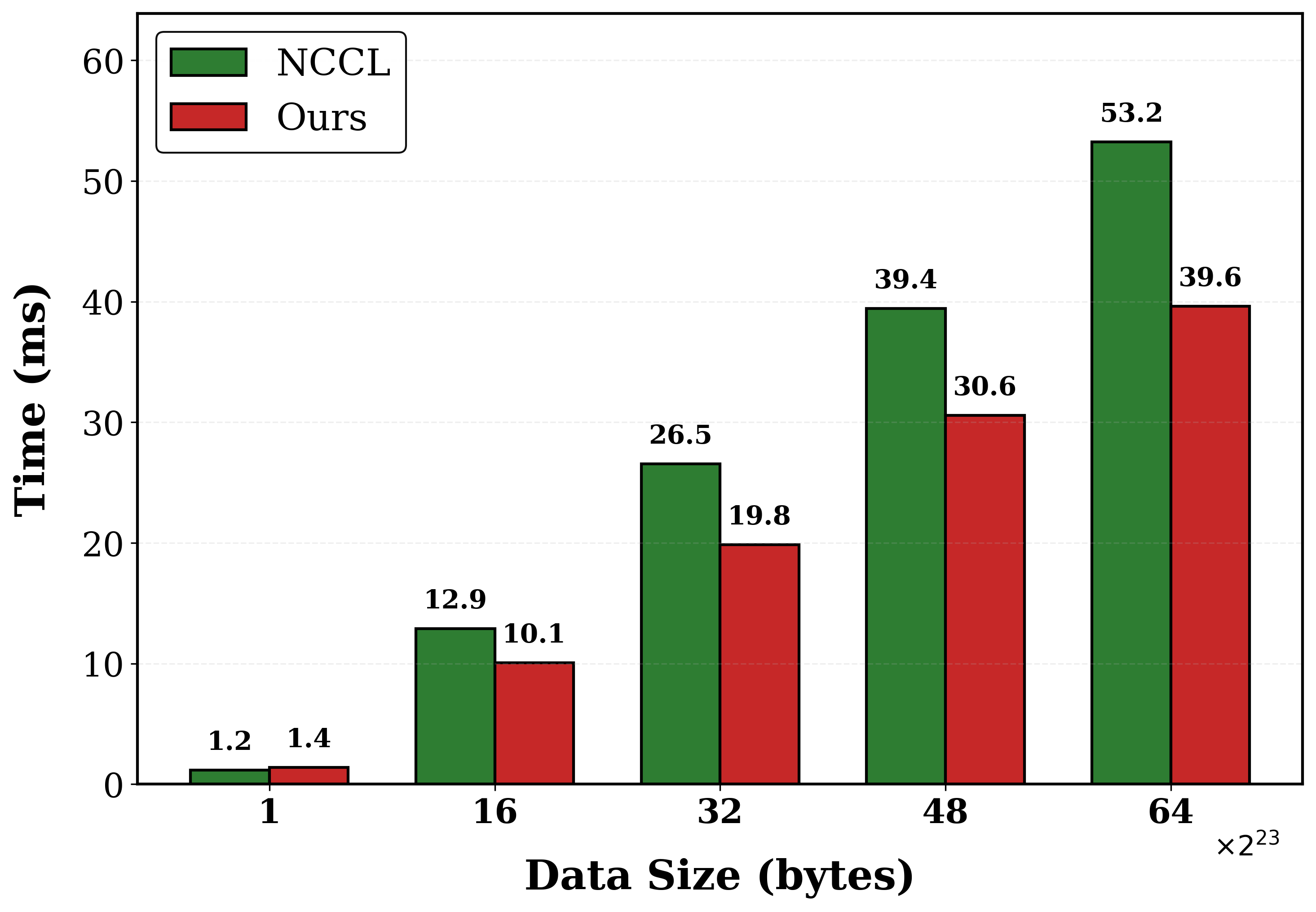}
		\caption{All-Gather time with varied communication volume}
		\label{fig:ag_vol}
	\end{subfigure}
    
    	 \begin{subfigure}[b]{0.23\textwidth}
		\centering
		\includegraphics[width=\textwidth]{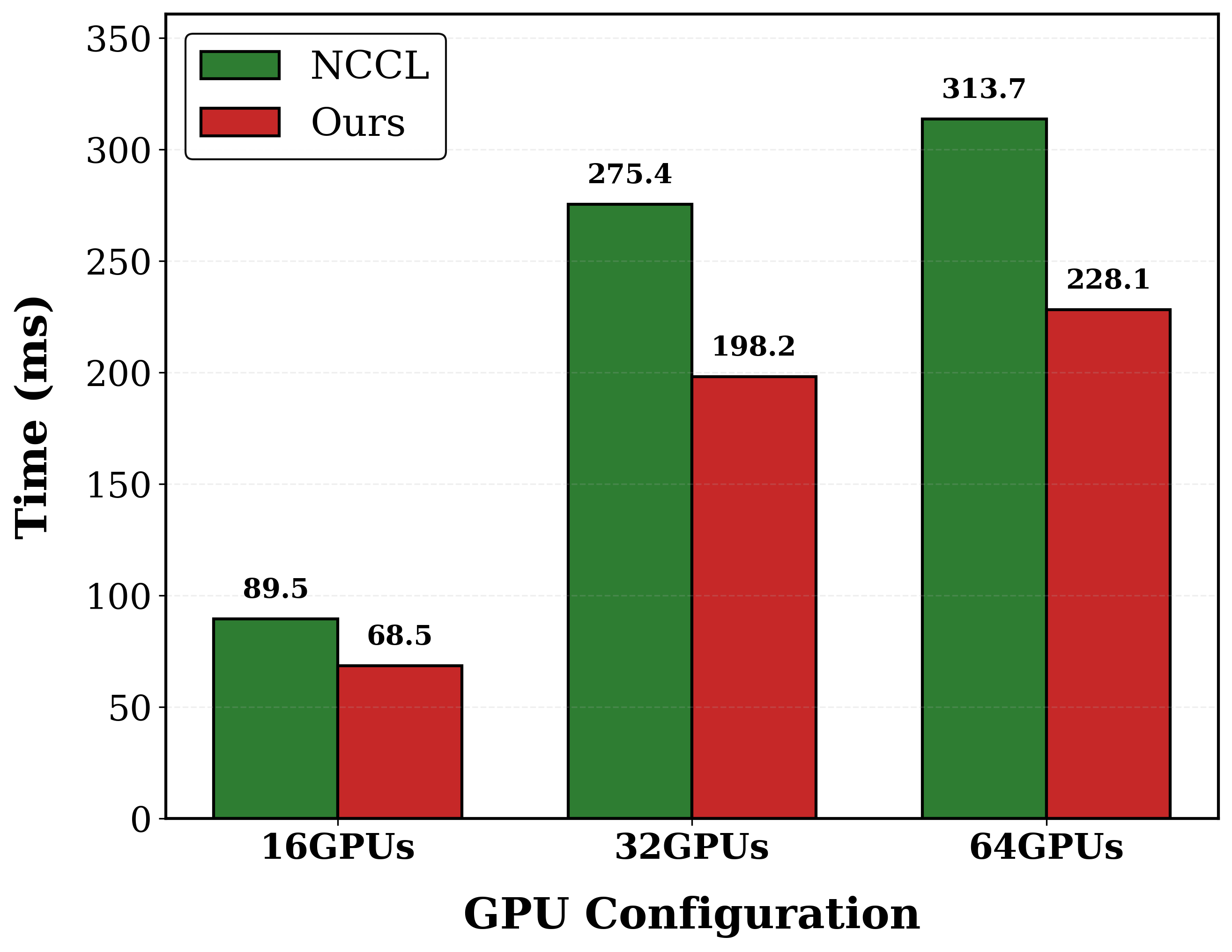}
		\caption{All-to-All time with varied number of GPUs}
		\label{fig:a2a_gpu}
	\end{subfigure}
      \begin{subfigure}[b]{0.23\textwidth}
		\centering
		\includegraphics[width=\textwidth]{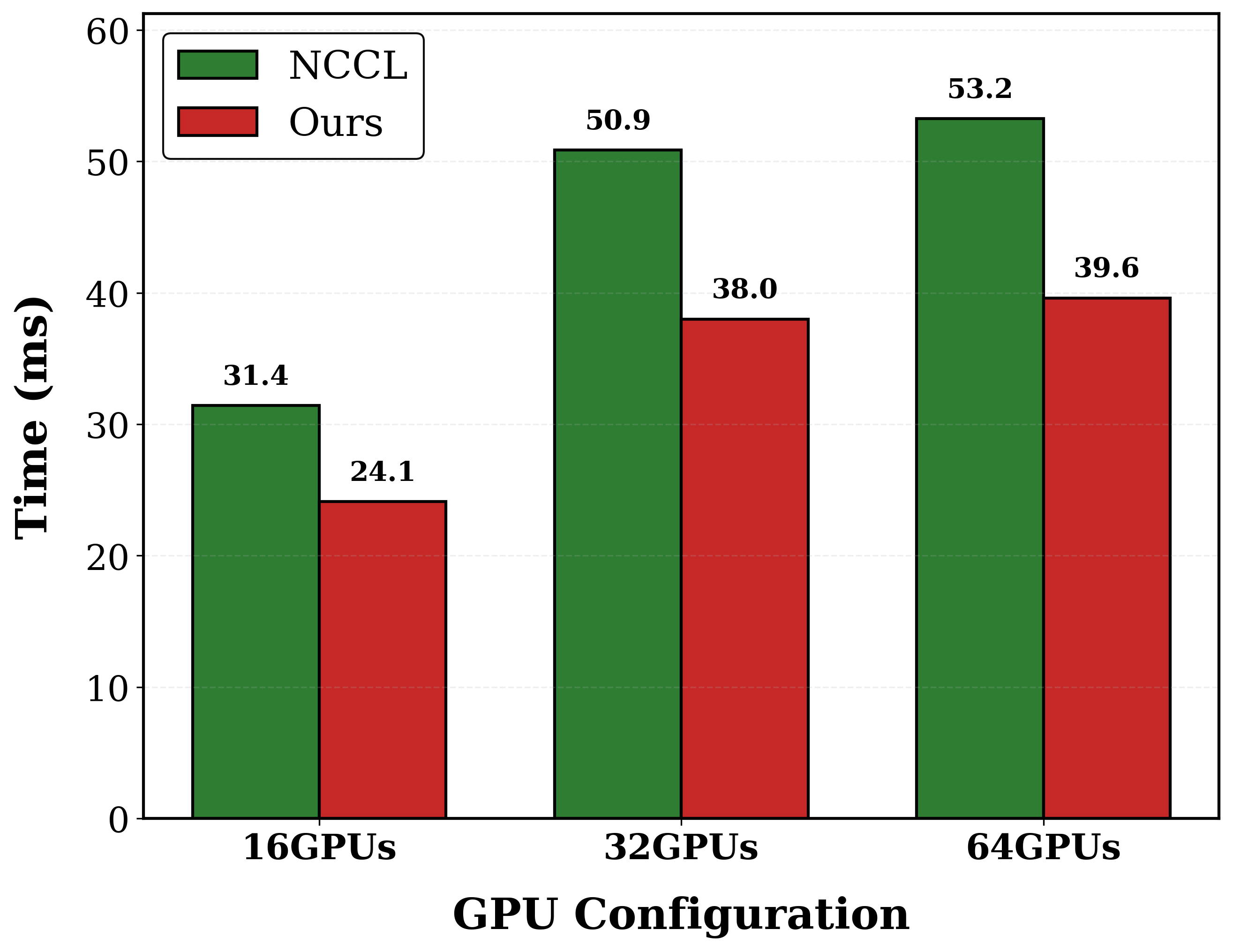}
		\caption{All-Gather time with varied number of GPUs}
		\label{fig:ag_gpu}
	\end{subfigure}
	\caption{All-to-All and All-Gather time of \modelname{} and NCCL on different configuration}
	\label{fig:compare_nccl}
\end{figure}
\textbf{Collective Performance Comparison with NCCL. }
As shown in Fig.~\ref{fig:compare_nccl}, we compare the All‑Gather and All‑to‑All times of \modelname{} against NCCL across different GPU counts and communication volumes. When the data volume is small, bandwidth cannot be fully utilized, and our \modelname{} shows very close performance to NCCL. However, as the data volume increases (which is very common in LLM training) and bandwidth is better saturated, the improvement of our \modelname{} over NCCL becomes significant: on average, we achieve $1.36\times$ speedup for All‑to‑All and $1.31\times$ for All‑Gather.

\subsection{Ablation Study}
We conduct ablation studies in this section to evaluate the methods introduced in \S\ref{sec:imbalance} and \S\ref{sec:switch}.
\begin{figure}[!t]
	\centering
		\includegraphics[width=0.48\textwidth]{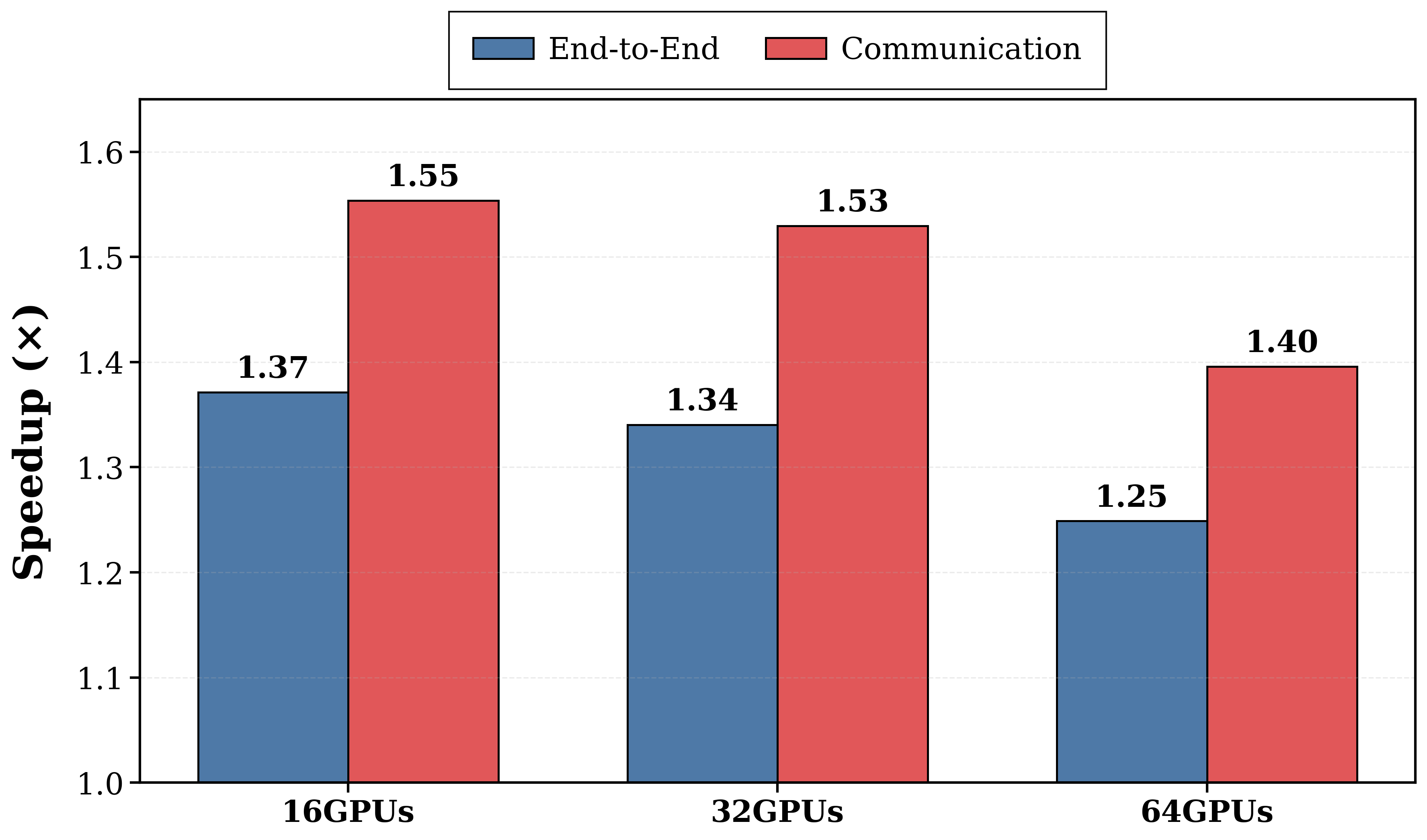}
	\caption{End-to-end and All-to-All communication speedups of Zipped All-to-All Design-2 over Design-1 on Qwen3-MoE.}
	\label{fig:IMBAL}
\end{figure}

\textbf{Effect of the Two All‑to‑All Designs.}
To validate the practical value of the two All‑to‑All designs introduced in \S\ref{sec:imbalance}, we conduct experiments on Qwen3-MoE using the same configuration as in \S\ref{sec:e2e_compare}. Results shown in Fig.~\ref{fig:IMBAL} demonstrate that the Zipped All‑to‑All Design‑1 (Fig.~\ref{fig:a2a1}) completely loses the benefit of communication compression under imbalanced data distributions, even increasing communication time and slowing down end‑to‑end training. In contrast, Zipped All‑to‑All Design‑2 (Fig.~\ref{fig:a2a2}) successfully retains the advantage of compressed communication, improving end‑to‑end performance by 1.25$\times$–1.37$\times$ and communication speed by 1.40$\times$–1.55$\times$ relative to Design‑1.
\begin{figure}[!t]
	\centering
		\includegraphics[width=0.48\textwidth]{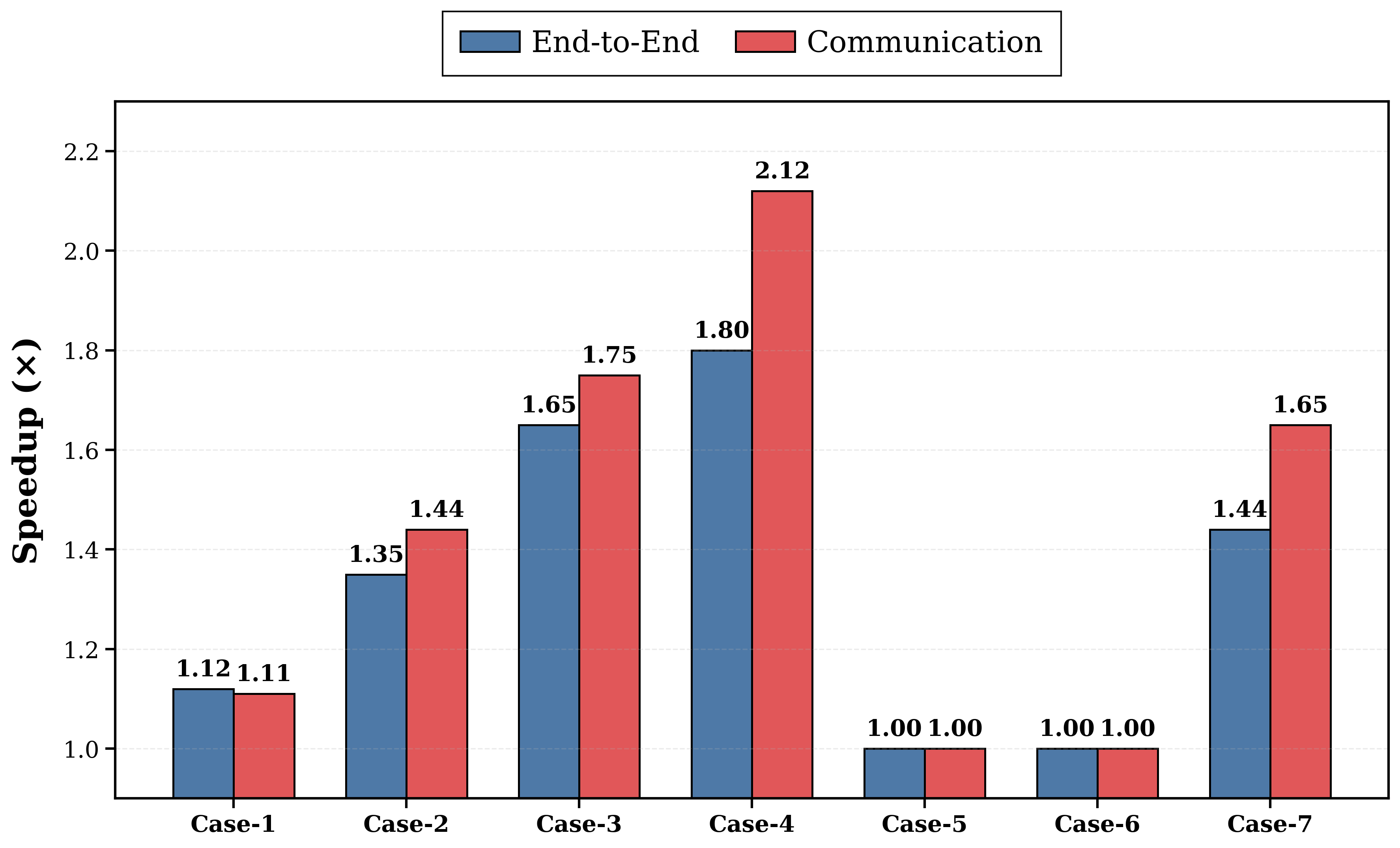}
	\caption{End-to-end and All-to-All communication speedups of our \modelname{} with the adaptive switcher compared to the base implementation without the adaptive switcher across seven scenarios.}
	\label{fig:SW}
\end{figure}

\textbf{Effect of the Adaptive Switcher for Reduce‑Scatter. }
To evaluate the adaptive switcher proposed in \S\ref{sec:switch}, we conduct Llama3-8B training experiments on TorchTitan with seven different hardware configurations:
(1) 1 node, 8 GPUs; (2) 2 nodes, 16 GPUs; (3) 4 nodes, 32 GPUs; (4) 8 nodes, 64 GPUs;  
(5) 8 nodes, 8 GPUs (1 GPU per node); (6) 8 nodes, 16 GPUs (2 GPUs per node); (7) 8 nodes, 32 GPUs (4 GPUs per node).

These configurations represent different efficiencies between zipped and native Reduce‑Scatter. We compare end-to-end training time and Reduce‑Scatter latency with and without the adaptive switcher. As shown in Fig.~\ref{fig:SW}, the switcher consistently delivers equal or better performance: for configurations (6) and (7), it matches the baseline, while for all others, it achieves measurable speedups, providing an average improvement of 1.39$\times$.

\begin{figure}[!t]
	\centering
 \begin{subfigure}[b]{0.1\textwidth}
		\centering
		\includegraphics[width=\textwidth]{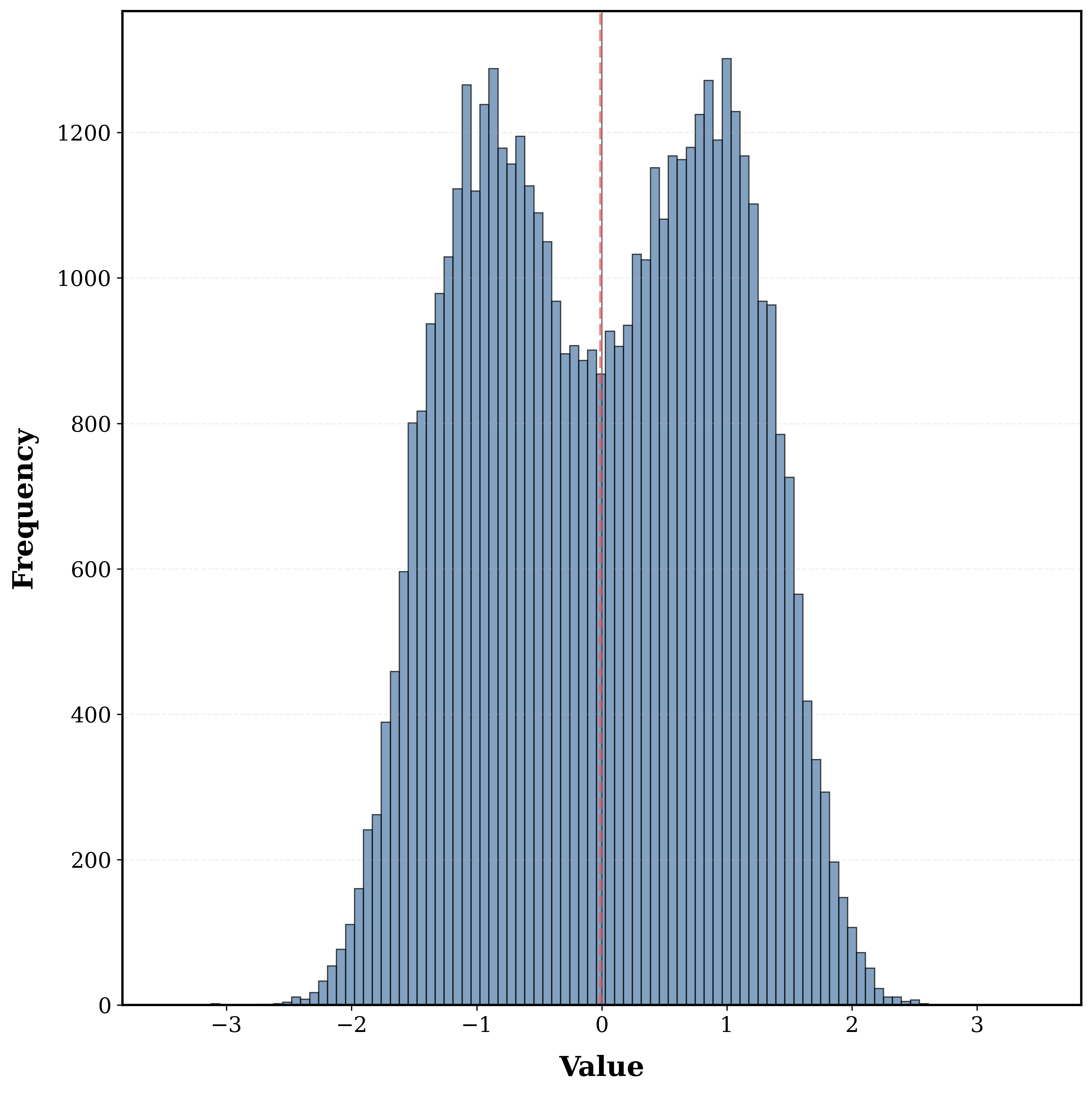}
		\caption{I-1}
	\end{subfigure}
     \begin{subfigure}[b]{0.1\textwidth}
		\centering
		\includegraphics[width=\textwidth]{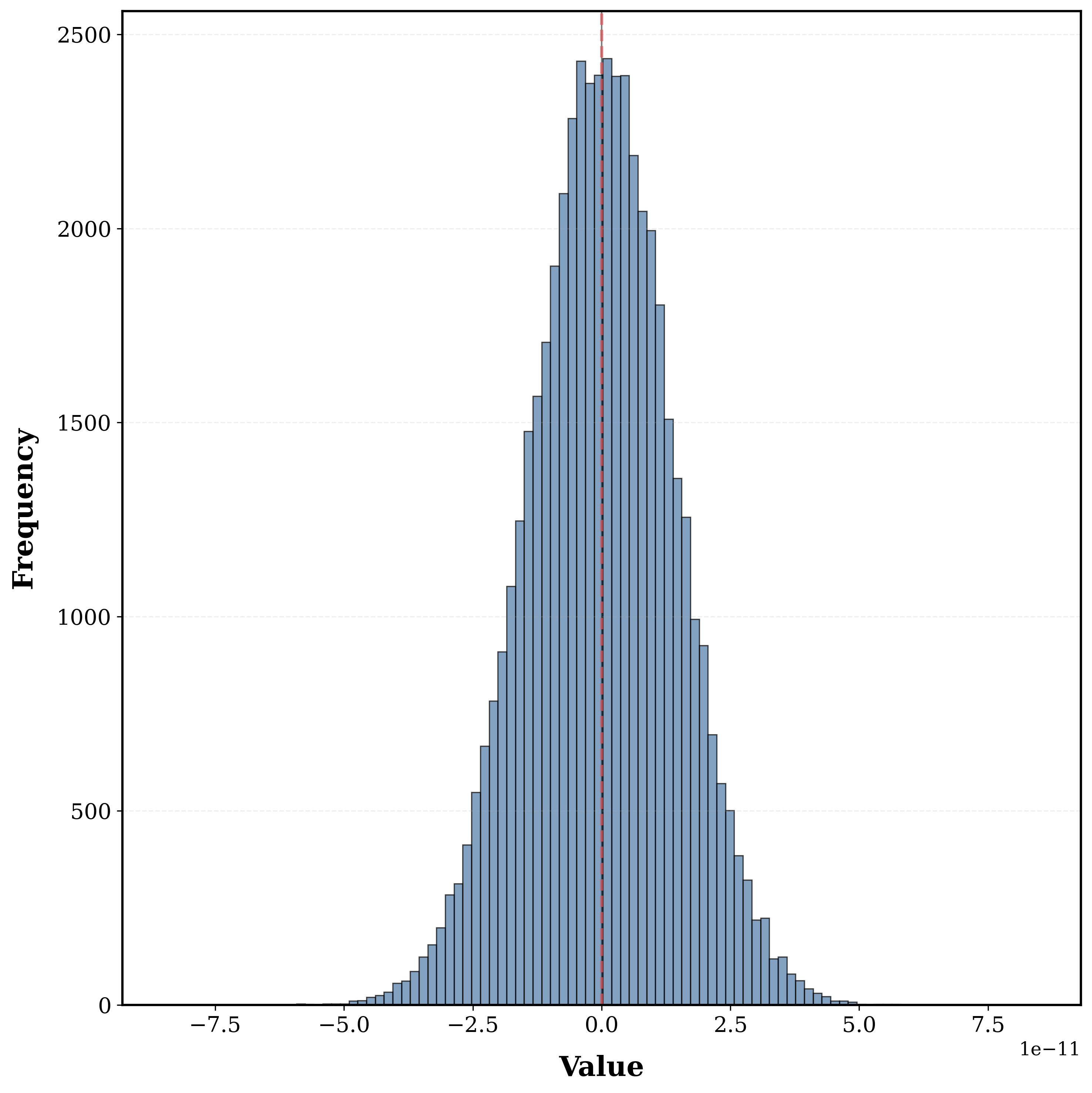}
		\caption{IG-1}
	\end{subfigure}
    	 \begin{subfigure}[b]{0.1\textwidth}
		\centering
		\includegraphics[width=\textwidth]{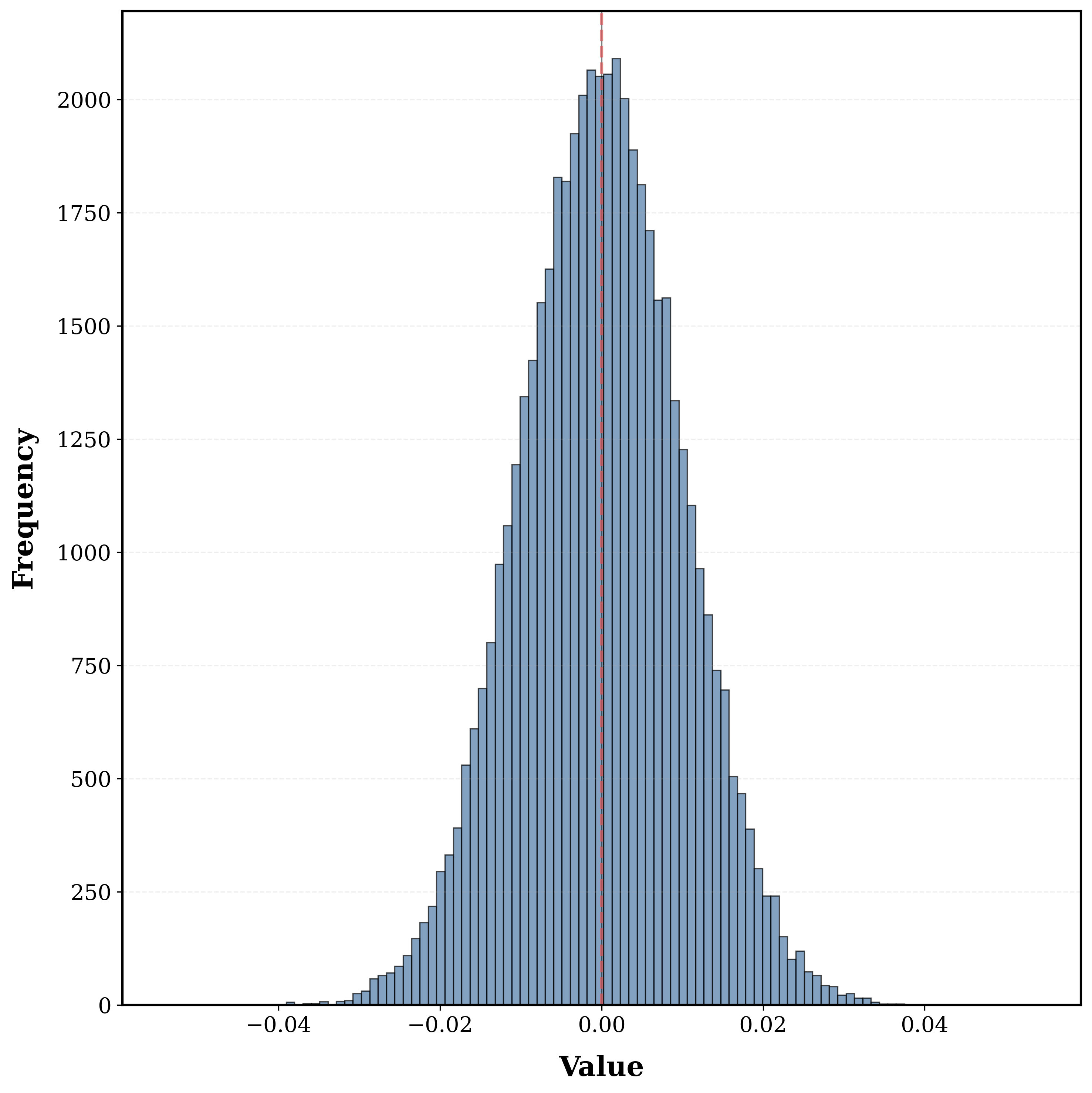}
		\caption{W-1}
	\end{subfigure}
      \begin{subfigure}[b]{0.1\textwidth}
		\centering
		\includegraphics[width=\textwidth]{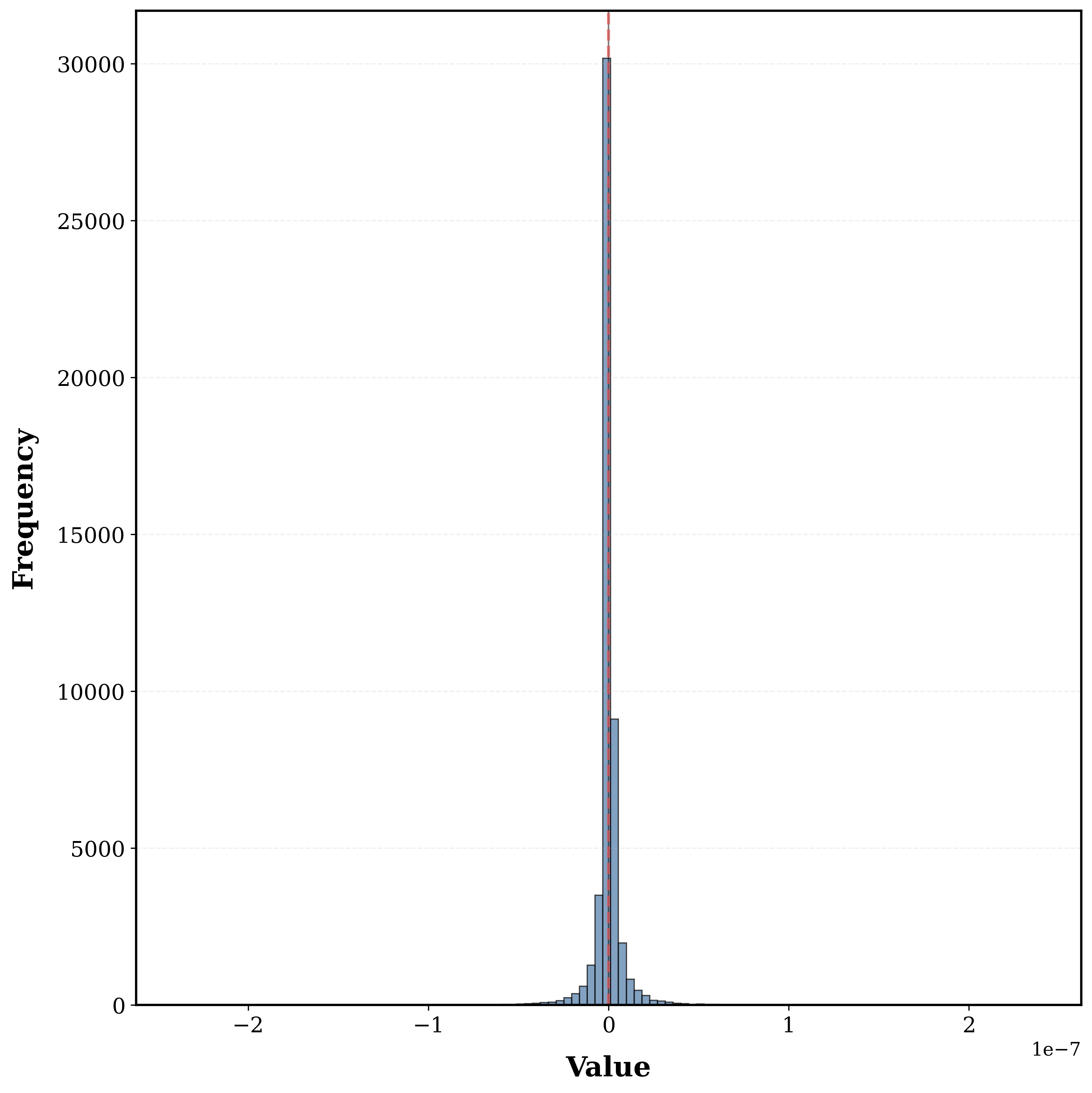}
		\caption{WG-1}
	\end{subfigure}

     \begin{subfigure}[b]{0.1\textwidth}
		\centering
		\includegraphics[width=\textwidth]{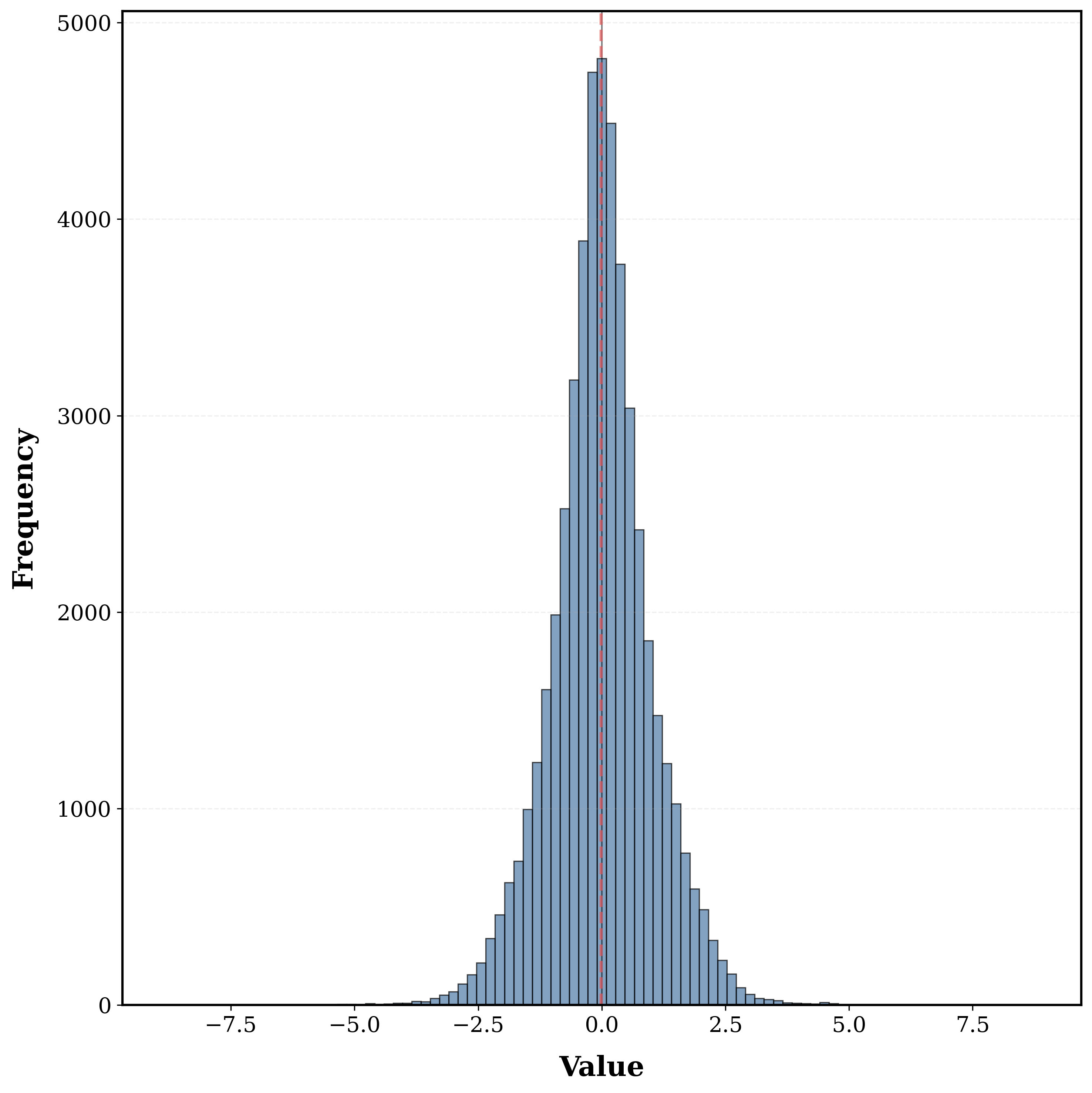}
		\caption{I-2}
	\end{subfigure}
     \begin{subfigure}[b]{0.1\textwidth}
		\centering
		\includegraphics[width=\textwidth]{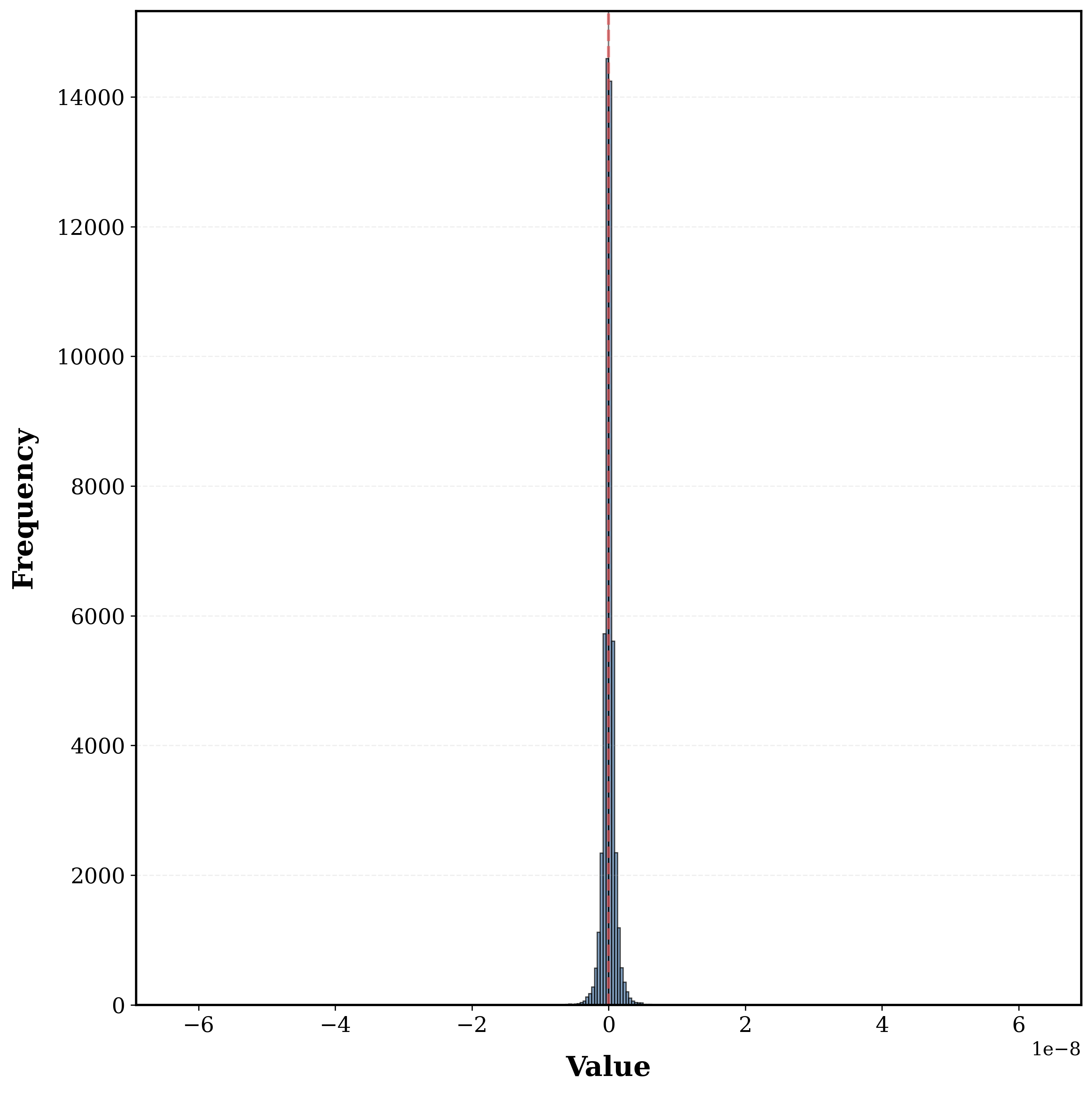}
		\caption{IG-2}
	\end{subfigure}
    	 \begin{subfigure}[b]{0.1\textwidth}
		\centering
		\includegraphics[width=\textwidth]{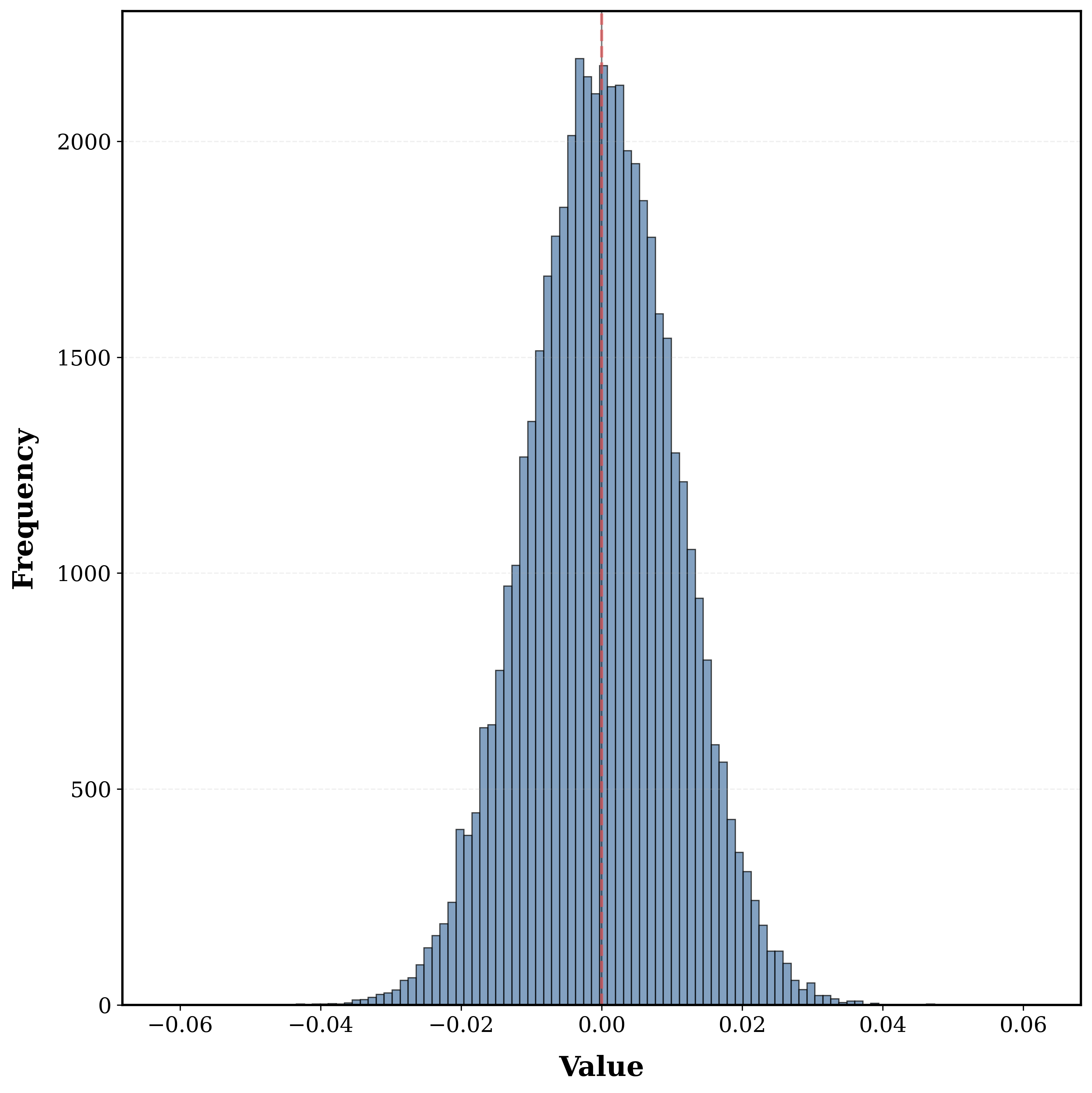}
		\caption{W-2}
	\end{subfigure}
      \begin{subfigure}[b]{0.1\textwidth}
		\centering
		\includegraphics[width=\textwidth]{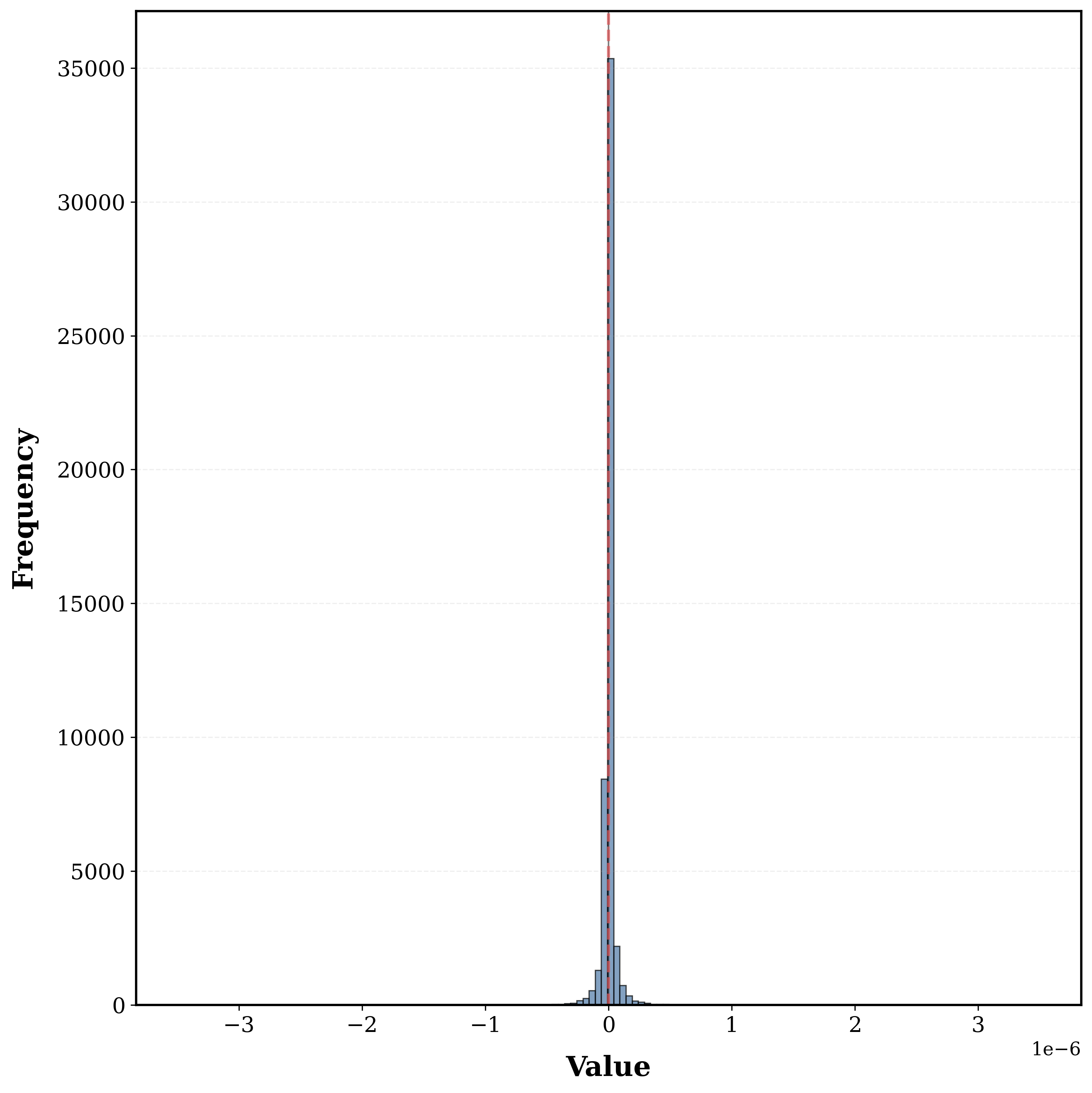}
		\caption{WG-2}
	\end{subfigure}

     \begin{subfigure}[b]{0.1\textwidth}
		\centering
		\includegraphics[width=\textwidth]{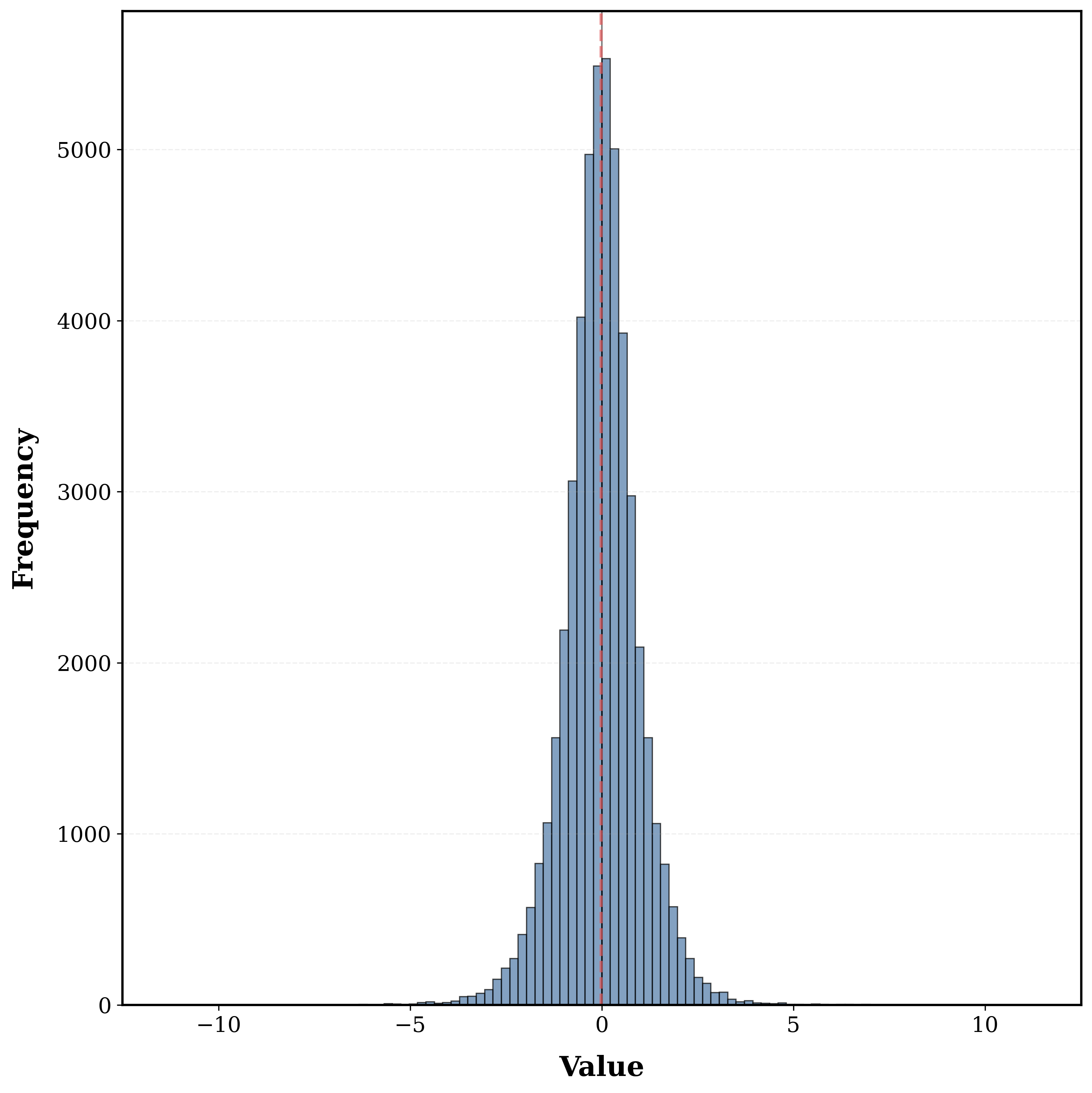}
		\caption{I-3}
	\end{subfigure}
     \begin{subfigure}[b]{0.1\textwidth}
		\centering
		\includegraphics[width=\textwidth]{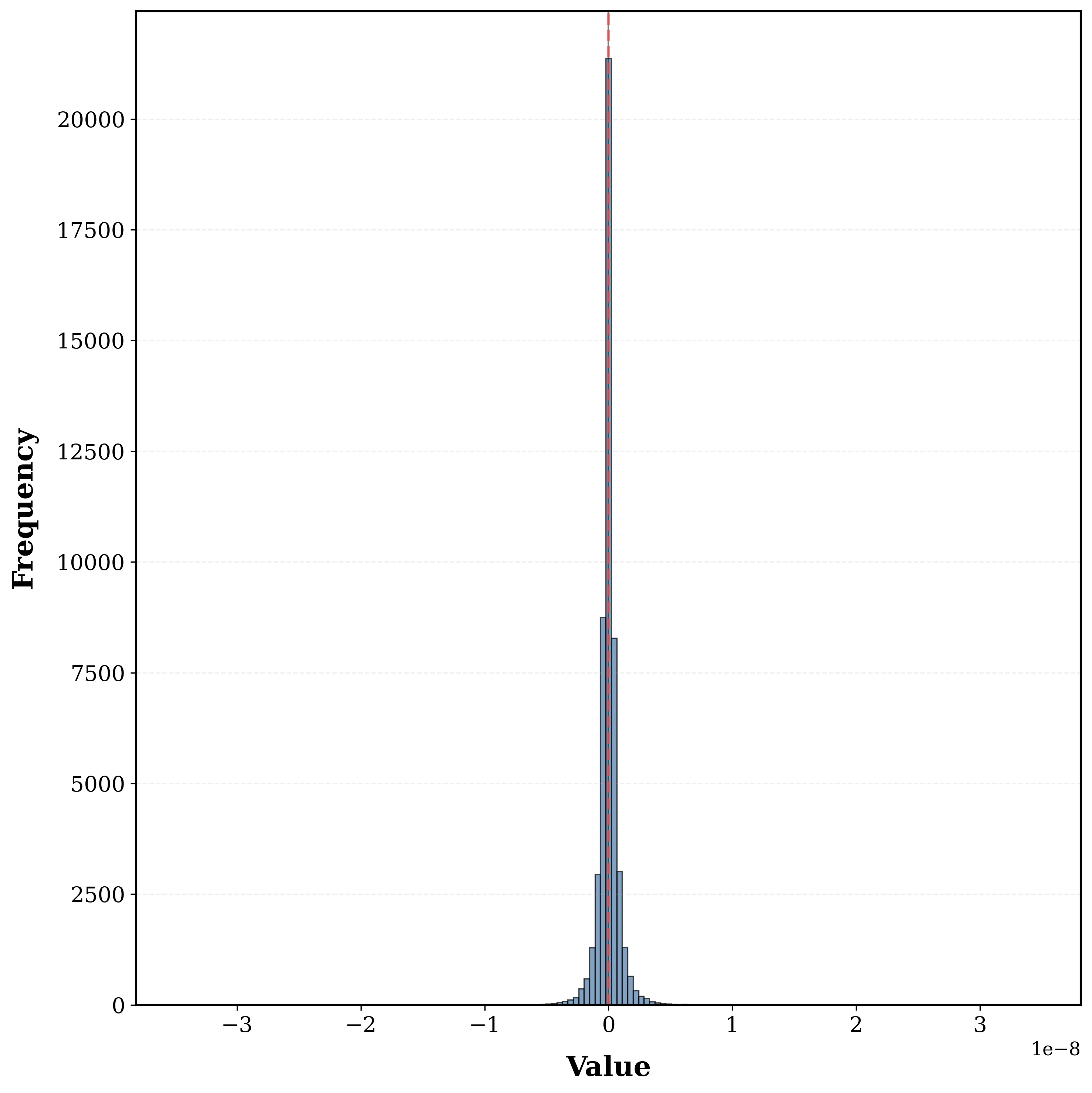}
		\caption{IG-3}
	\end{subfigure}
    	 \begin{subfigure}[b]{0.1\textwidth}
		\centering
		\includegraphics[width=\textwidth]{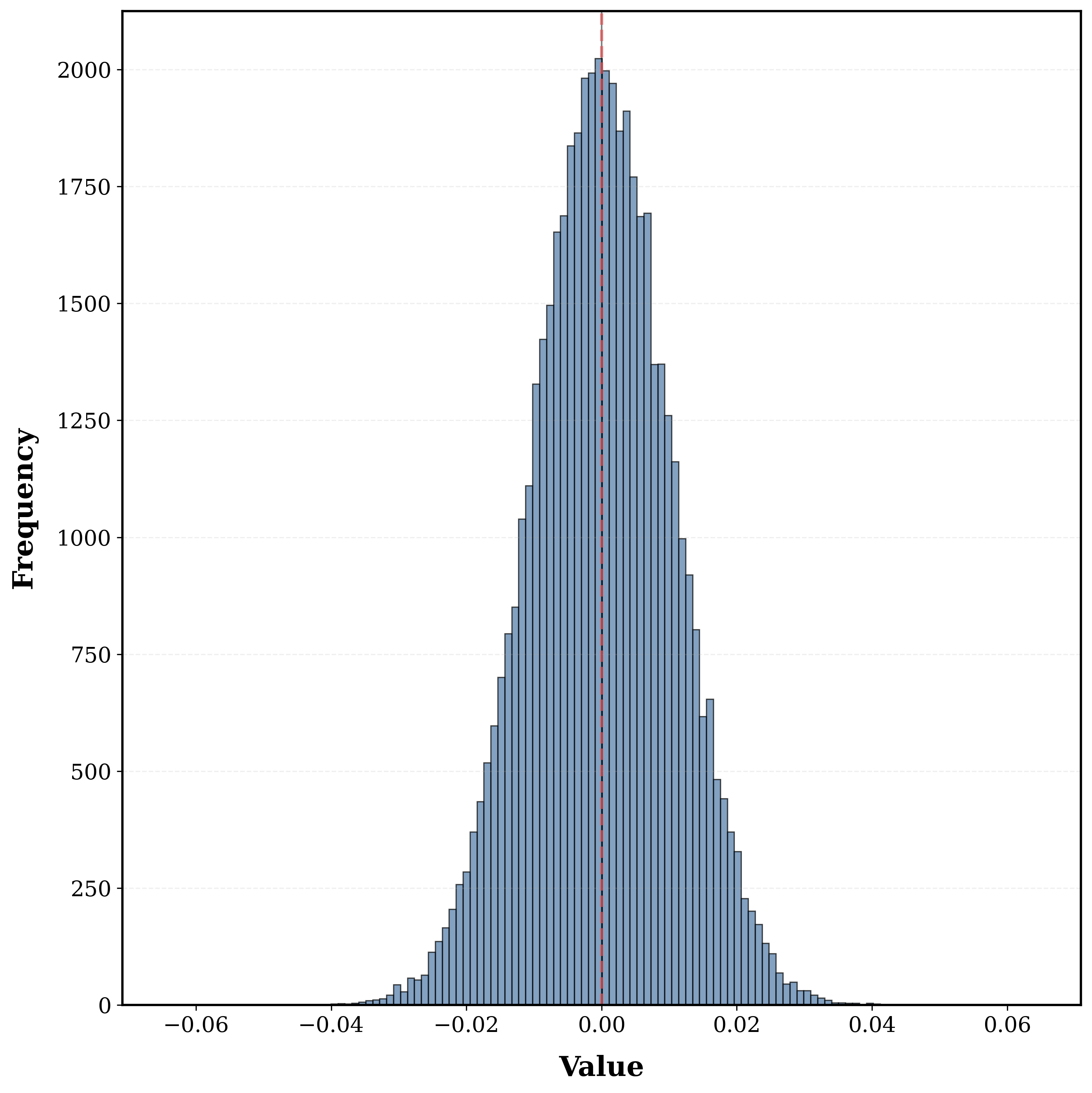}
		\caption{W-3}
	\end{subfigure}
      \begin{subfigure}[b]{0.1\textwidth}
		\centering
		\includegraphics[width=\textwidth]{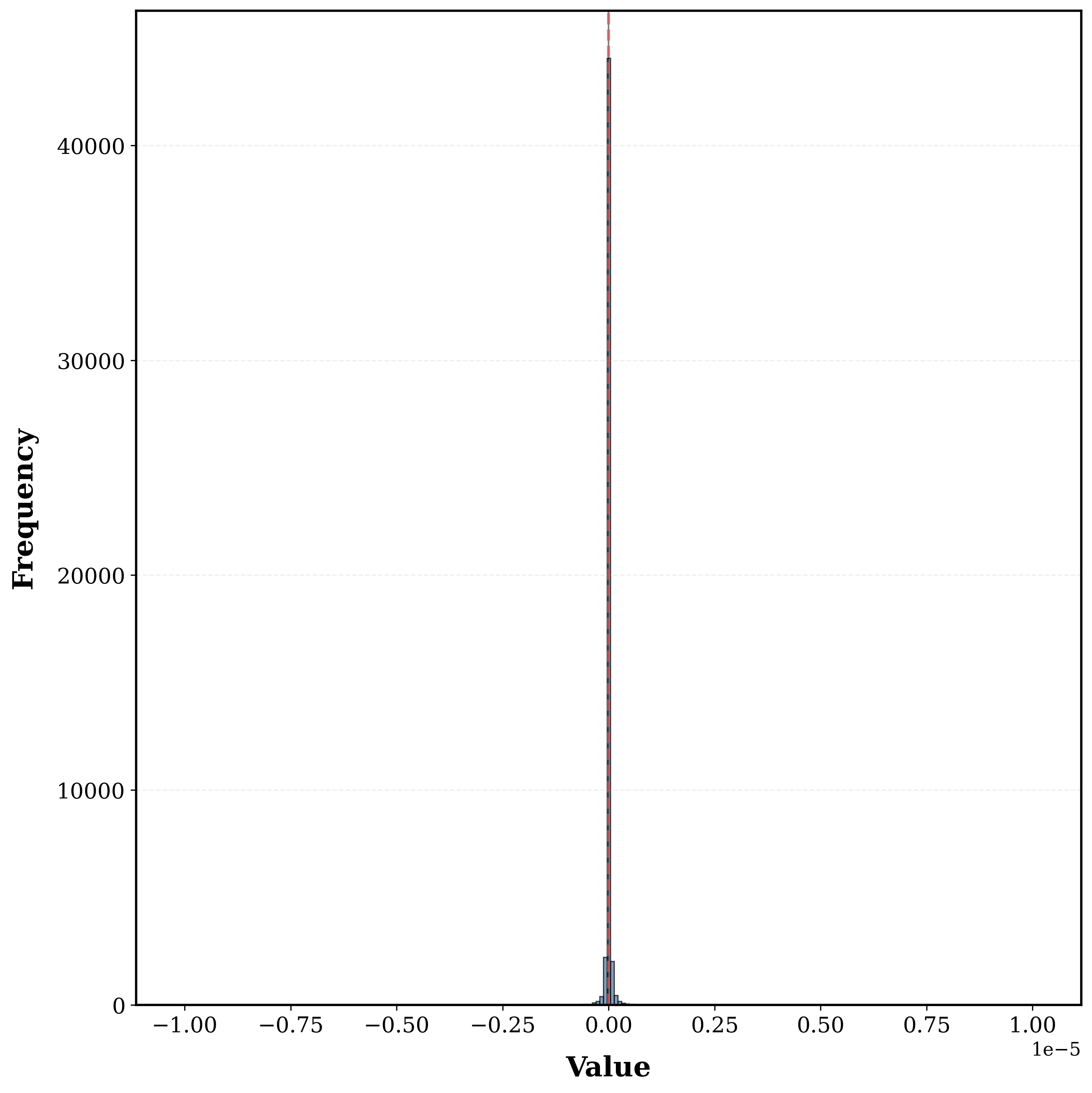}
		\caption{WG-3}
	\end{subfigure}
	\caption{Data distribution of the activation, weight and gradient in Llama3-8B training.}
	\label{fig:norm}
\end{figure}

\subsection{Data Normality in LLM Training}\label{subsec:norm}
To validate our assumption that the data follows a normal or near‑Gaussian distribution, we conducted additional experiments by training a Llama3‑8B model on 2B tokens from OpenWebText~\cite{peterson2019open}. We collected the numerical distributions of the input activations (I), weights (W), and their corresponding gradients (IG, WG) from the first matrix multiplication in the FFN module of the first transformer layer at three training checkpoints: after 0.002B (-1), 1B (-2), and 2B (-3) tokens. As illustrated in Fig.~\ref{fig:norm}, all three distributions: activations, weights, and gradients consistently exhibit normal or near‑Gaussian shapes throughout the training process. While I-1 as shown in Fig.~\ref{fig:norm}(a) demonstrates a particular shape, it becomes near-Gaussian shapes afterward as shown in Fig.~\ref{fig:norm}(e)(i).
\section{Related Works}
\textbf{Lossless Compression.} LMC~\cite{waddington2025lossless} and ZipNN~\cite{zipnn} employ Huffman coding~\cite{huffman2007method} to compress model checkpoints for efficient storage and transmission, whereas NeuZip~\cite{neuzip} targets runtime memory reduction. DietGPU~\cite{dietgpu} and nvCOMP~\cite{NVIDIA_nvcomp_2025} specialize in GPU‑accelerated lossless compression and decompression. UCCL-Zip~\cite{2026arXiv260417172M}, DFloat11~\cite{dfloat11} and ZipServ~\cite{zipserv} use compression to accelerate inference or weights synchronization. Additionally, several hardware augmented lossless compression schemes have been proposed~\cite{kim2016bit, choukse2020buddy, ekman2005robust, zhao2015buri, pekhimenko2013linearly, choukse2018compresso,pekhimenko2012base}. However, a more generally applicable software‑only approach for lossless compression in communication remains underexplored.

\textbf{Communication Optimization.} We can roughly divide current communication optimization research into three categories. The first category centers on scheduling~\cite{shi2023pipemoe,lin2025mast,pan2025fsmoe,jiang2024lancet,wang2025spmoe}, where computation is overlapped with communication to hide communication latency. 
% For example, FSMoE~\cite{pan2025fsmoe} further pipelines different communication operations across distinct resources. Lancet~\cite{jiang2024lancet} and SP-MoE~\cite{wang2025spmoe} leverage attention computation to overlap with All-to-All communication. 
The second category focuses on the network topology~\cite{lin2025hiermoe,blink,liu2024deepseek}. 
% HierMoE~\cite{lin2025hiermoe} considers network topology and the duplication problem in MoE training, thereby reducing communication overhead. Blink~\cite{blink} models heterogeneous hardware topologies as graphs to capture bandwidth and latency, enabling topology-aware collective communication. 
The third category adopts communication compression, primarily utilizing lossy techniques such as quantization~\cite{shi2024schemoe} or sparsification~\cite{Jin_Wang_Zhu_Zhan_Bai_Zhang_Ming_Li_2025}. While effective in reducing data volume, these methods introduce irreversible information loss. 
% ScheMoE~\cite{shi2024schemoe} quantizes data before transmission to cut communication overhead while preserving training accuracy. BigMac~\cite{Jin_Wang_Zhu_Zhan_Bai_Zhang_Ming_Li_2025} applies a projection to shrink the activation hidden dimension before communication. Ladder-Residual~\cite{zhang2025ladderresidual} redesigns the model architecture to decouple communication from computation. 
Different from previous works, we focus on a more generally practical but rarely explored
method for communication optimization: lossless compression communication.

\section{Conclusion}
In this work, we present \modelname{}, a library of lossless compressed communication collectives designed to alleviate the communication bottleneck in distributed LLM training. Our approach effectively bridges the gap between the theoretical compression potential of near Gaussian LLM tensors and practical GPU performance through three core innovations: (1) a theoretically grounded exponent coding scheme that eliminates costly online statistics collection, (2) GPU-optimized compression and decompression kernels with communication-aware data layout, carefully design memory access patterns, and pipelining, and (3) adaptive communication strategies for All‑to‑All operations in MoE training that handle imbalanced expert computation loads, along with a dynamic switcher for Reduce‑Scatter that adapts to underlying system characteristics.
Through extensive evaluation on a 64‑GPU cluster using both MoE and dense transformer models, we demonstrate that \modelname{} significantly reduces communication latency, achieving up to 1.35$\times$ faster communication and 1.18$\times$ end‑to‑end training speedups on Megatron-LM and TorchTitan, while preserving bit‑exact correctness.
\bibliographystyle{ACM-Reference-Format}
\bibliography{reference}

\end{document}